\documentclass[aps,twocolumn,groupedaddress,showpacs]{revtex4-1}
\usepackage{graphicx,amsmath,amssymb}
\usepackage{array,bigstrut}
\usepackage[colorlinks]{hyperref}

\usepackage{ntheorem}
\theoremstyle{break}
\theorembodyfont{\upshape}
\newtheorem{theorem}{Theorem}
\newtheorem{property}{Property}

\begin{document}


\title{Complementary-multiphase quantum search for all numbers of target items}

\author{Tan Li}

\author{Wan-Su Bao}
\email{2010thzz@sina.com}

\author{He-Liang Huang}

\author{Feng-Guang Li}

\author{Xiang-Qun Fu}

\author{Shuo Zhang}

\author{Chu Guo}

\author{Yu-Tao Du}

\author{Xiang Wang}

\author{Jie Lin}

\affiliation{Henan Key Laboratory of Quantum Information and Cryptography, Zhengzhou Information Science and Technology Institute, Zhengzhou, Henan 450001, China}
\affiliation{Synergetic Innovation Center of Quantum Information and Quantum Physics, University of Science and Technology of China, Hefei, Anhui 230026, China}


\begin{abstract}
Grover's algorithm achieves a quadratic speedup over classical algorithms,
but it is considered necessary to know the value of $\lambda$ exactly
[Phys. Rev. Lett. 95, 150501 (2005); Phys. Rev. Lett. 113, 210501 (2014)], where
$\lambda$ is the fraction of target items in the database.
In this paper, we find out that
the Grover algorithm can actually apply to the case
where one can identify the range that $\lambda$ belongs to
from a given series of disjoint ranges.
However, Grover's algorithm still cannot maintain high success probability
when there exist multiple target items.
For this problem, we proposed a complementary-multiphase quantum search algorithm,
in which multiple phases complement each other
so that the overall high success probability can be maintained.
Compared to the existing algorithms, in the case defined above,
for the first time our algorithm achieves the following three goals simultaneously:
(1) the success probability can be no less than any given value between 0 and 1,
(2) the algorithm is applicable to the entire range of $\lambda$, and
(3) the number of iterations is almost the same as that of Grover's algorithm.
Especially compared to the optimal fixed-point algorithm
[Phys. Rev. Lett. 113, 210501 (2014)],
our algorithm uses fewer iterations to achieve success probability greater than 82.71\%,
e.g., when the minimum success probability is required to be 99.25\%,
the number of iterations can be reduced by 50\%.
\end{abstract}


\maketitle

\section{Introduction \label{sec:Introduction}}

For the unordered database search problem,
the Grover algorithm \cite{Grover1996,Grover1997} provides a quadratic improvement over classical search algorithms,
and has drawn considerable research attention.
However, it has been indicated that
``to perform optimally, they need precise knowledge of certain problem parameters,
e.g., the number of target states'' \cite{Grover2005},
and ``without knowing exactly how many marked items there are,
there is no knowing when to stop the iteration'' \cite{Yoder2014}.
In other words, the Grover algorithm is considered to be only applicable to the case,
denoted by Case-KPV (knowledge of precise value),
where the value of fraction of target items is precisely known.

In fact, the optimal number of iterations of Grover's algorithm is given by
(See p.~253 of Ref.~\cite{Nielson2000})
\begin{equation}
k_{G}=CI\Big(\frac{\pi}{4\arcsin\sqrt{\lambda}}-\frac{1}{2}\Big),\label{eq:optimal_k_of_Grover_CI}
\end{equation}
where $\lambda=M/N$ represents the fraction of target items,
$M$ is the number of target items in a database of $N$ items,
and $CI\left(x\right)$ returns the integer closest to $x$ and rounds halves down.
Simple algebra shows that
\begin{equation}
\!\!\!\!k_{G}=\!\begin{cases}
0, & \!\!\!{\rm if}\thinspace\lambda\in\!\big[\frac{1}{2},1\big)\equiv \!\varLambda_{G,0},\\
m, & \!\!\!{\rm if}\thinspace\lambda\in\!\big[\sin^{2}\!\frac{\pi}{4m+4},\sin^{2}\!\frac{\pi}{4m}\big)\equiv \!\varLambda_{G,m},m\!\ge\!1.
\end{cases}\label{eq:optimal_k_of_Grover_step}
\end{equation}
Then the optimal number of iterations $k_{G}$ can be determined provided one can identify
which of the given ranges $\{\varLambda_{G,m}:m\ge0\}$ that $\lambda$ belongs to.

\begin{figure}[tb]
\begin{centering}
\includegraphics[width=5.5cm]{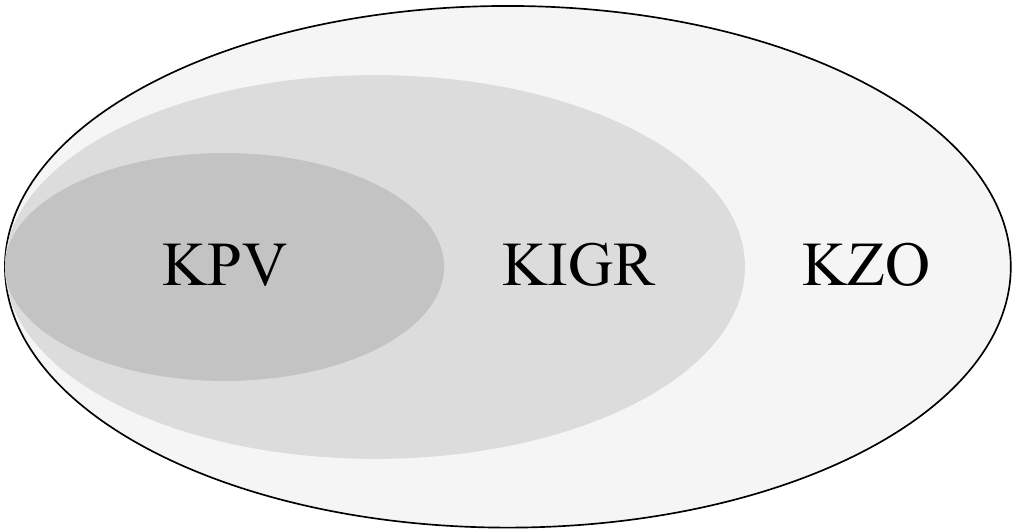}
\par\end{centering}

\protect\caption{Inclusion relationships among Cases-KPV, KIGR and KZO.
In Case-KPV, $\lambda$ is precisely known.
In Case-KIGR, the range that $\lambda$ belongs to can be identified
from a given series of disjoint ranges.
In Case-KZO, one knows that $0<\lambda<1$.
\label{fig:relationships_among_Cases}}
\end{figure}

Consequently, we confirm that Grover's algorithm is applicable to the case,
denoted by Case-KIGR (knowledge of identifiability in given ranges),
where one can identify the range that $\lambda$ belongs to
from a given series of disjoint ranges of $\lambda$.
As illustrated in Fig.~\ref{fig:relationships_among_Cases},
Case-KIGR includes Case-KPV.
For example, if $\lambda$ is not precisely known,
but knowing that $\lambda\in\left[0.2,0.5\right)$, then
from $\{\varLambda_{G,m}\}$ we can identify
$\lambda\in\varLambda_{G,1}\approx\left[0.1464,0.5\right)$, thus $k_{G}=1$, which
shows that the Grover algorithm is still applicable.
Note that, the given ranges in the definition of Case-KIGR can be different in different algorithms.

The Grover algorithm has been proven optimal \cite{Bennett1997,Boyer1998,Zalka1999,Grover2005a}.
However, the minimum success probability of Grover's algorithm is only 50\%.
For this problem, quantum amplitude amplification \cite{Brassard1997a,Grover1998,Brassard1998,Brassard2002}
with arbitrary phases has been developed, as well as the phase matching methods \cite{Long1999,Hyer2000,Long2001a,Long2002,Li2002,Li2007}.
Furthermore, many generalizations and modifications of Grover's algorithm have been proposed \cite{Biron1999,Younes2004,Giri2017,Byrnes2018}.

There is a natural problem here, i.e., in Case-KIGR, is there such an algorithm that preserves the advantages
(i.e., the algorithm applies to the entire range of $\lambda$
and the number of iterations remains almost the same as the Grover algorithm),
and also overcomes the success probability problem of Grover's algorithm?

First, the 100\%-success probability algorithms \cite{Brassard2002,Chi1999,Long2001,Toyama2013,Liu2014}
which can complete searching with certainty,
are only applicable to Case-KPV,
because precise knowledge of $\lambda$ is necessary to
determine the phase of the algorithm.

Next, for the fixed-phase algorithms \cite{Younes2013,Zhong2008,Li2012},
the phase is first fixed to a certain value, being independent of $\lambda$,
then the optimal number of iterations \cite{Li2012} is specified by
\begin{equation}
k=\big\lfloor
{\pi}/4\cdot{\arcsin^{-1}\big[\sqrt{\lambda}\sin\big(\phi/2\big)\big]}
\big\rfloor,\label{eq:optimal_k_of_Li2012}
\end{equation}
where $\left\lfloor \cdot\right\rfloor $ is the floor function.
$k$ is a step function of $\lambda$,
therefore these algorithms can also apply to Case-KIGR.
However, as we have seen in Eq.~(\ref{eq:optimal_k_of_Li2012}),
more iterations are required than the Grover algorithm, when $\phi\neq\pi$.

Then, in the original matched-multiphase algorithms \cite{Toyama2008,Toyama2009},
phases are obtained by means of numerical fitting,
and only restricted ranges rather than $\left(0,1\right)$ can be covered.
Later by Yoder et al., Ref.~\cite{Yoder2014} provides phases in the analytical form,
and thus achieves the fixed point property, which makes the algorithm
overcome the souffl\'e problem \cite{Brassard1997} and apply to the most general case,
denoted by Case-KZO (knowledge of between zero and one),
where one knows that $0 < \lambda < 1$.
However, for the purpose of success probability no less than $1-\delta^{2}$,
the required number of iterations satisfies
\begin{equation}
k\ge\frac{\log\left(2/\delta\right)}{2\sqrt{\lambda}}-\frac{1}{2},\label{eq:optimal_k_of_yoder2014}
\end{equation}
which would be much larger than
that of Grover's algorithm for small enough $\delta$.
Especially when $\delta=0$, Yoder's algorithm becomes
the original fixed-point algorithm \cite{Grover2005},
and loses the quadratic speedup.
In addition, to Case-KZO, the trial-and-error algorithm \cite{Boyer1998,Brassard2002} is also applicable, which repeats the Grover algorithm with varying number of iterations.
However, the upper bound of expected number of iterations is about 10.19 times of Grover's algorithm, when $0<\lambda\le 3/4$ \cite{Boyer1998}.

Finally, in Ref.~\cite{Li2014} we presented a complementary-multiphase algorithm
that divides the range $\left[{1}/{4},1\right)$ into a series of small ranges,
each of which is specified a phase and number of iterations.
Thus the algorithm works well in Case-KIGR.
With one iteration, the success probability
can be no less than any $P_{cri}\in\left(0,1\right)$.
However, the success probability decreases to zero for $\lambda<{1}/{4}$,
which indicates that this method is no longer meaningful.

Therefore, in Case-KIGR, there is currently no algorithm that
overcomes the success probability problem and also
preserves the advantages of Grover's algorithm.
In this paper we expect to design
a complementary-multiphase quantum search algorithm with general iterations,
taking into account the success probability, the applicable range of $\lambda$ and
the number of iterations simultaneously,
and confirm that the multiphase-complementing method can actually apply to
the entire range of $\lambda$, by casting off the limitation of
$k=1$ in Ref.~\cite{Li2014}.

The paper is organized as follows.
Section~\ref{sec:Quantum_amplitude_amplification} provides an introduction to
the quantum amplitude amplification algorithm,
as well as the derivation of all local maximum points of
the success probability after $k$ iterations.
Section~\ref{sec:Generalized_complementary_multip} describes
the model of complementary-multiphase algorithm with general iterations,
and also the selection method of optimal parameters. Section~\ref{sec:Analysis_of_performance} gives an analysis of
the success probability and the number of iterations.
Section~\ref{sec:Discussions} summarizes the comparisons between
the algorithm in this paper and the existing algorithms,
followed by a brief conclusion in Section~\ref{sec:Conclusion}.

\section{Quantum amplitude amplification revisited \label{sec:Quantum_amplitude_amplification}}

Brassard et al. extended the phase inversions in the original Grover
algorithm \cite{Grover1996,Grover1997} to arbitrary rotations, and
obtained the quantum amplitude amplification algorithm \cite{Brassard1998,Brassard2002}.
The Grover iteration with arbitrary phases is given by
\begin{equation}
G\left(\phi,\varphi\right)=-HS_{0}^{\phi}HS_{f}^{\varphi}.\label{eq:arbitrary_phase_operation}
\end{equation}
Here $H$ is the Hadamard transform and
\begin{equation}
S_{f}^{\varphi}\left|x\right\rangle =\begin{cases}
e^{i\varphi}\left|x\right\rangle, & {\rm if}\thinspace f\left(x\right)=1,\\
\left|x\right\rangle, & {\rm if}\thinspace f\left(x\right)=0,
\end{cases}
\end{equation}
where $i=\sqrt{-1}$. Similarly, $S_{0}^{\phi}$ changes the phase of
zero state $\left|0\right\rangle $ by a factor of $\phi$. $S_{f}^{\varphi}$
and $S_{0}^{\phi}$ can be expressed as \cite{Long1999}
\begin{eqnarray}
S_{f}^{\varphi} & = & I-\left(-e^{i\varphi}+1\right)\sum_{x\in f^{-1}\left(1\right)}\left|x\right\rangle \left\langle x\right|,\\
S_{0}^{\phi} & = & I-\left(-e^{i\phi}+1\right)\left|0\right\rangle \left\langle 0\right|,
\end{eqnarray}
where $\varphi,\phi\in\left[0,2\pi\right)$, since $S_{f}^{\varphi}=S_{f}^{\varphi+2\pi}$
and $S_{0}^{\phi}=S_{0}^{\phi+2\pi}$.

The equal superposition of all target (nontarget) states can be denoted
by $\left|\alpha\right\rangle $ ($\left|\beta\right\rangle $), i.e.,
\begin{eqnarray}
\left|\alpha\right\rangle  & = & \frac{1}{\sqrt{M}}\sum\limits _{x\in f^{-1}\left(1\right)}\left|x\right\rangle ,\label{eq:target_state}\\
\left|\beta\right\rangle  & = & \frac{1}{\sqrt{N-M}}\sum\limits _{x\in f^{-1}\left(0\right)}\left|x\right\rangle ,\label{eq:nontarget_state}
\end{eqnarray}
where $N$ ($M$) is the number of all (target) items in the database, and by convention
$0<M<N$. Then, in the space spanned by $\left|\alpha\right\rangle $
and $\left|\beta\right\rangle $, the matrix representation of $G$
operator is
\begin{equation}
G\!=\!\!\left[\!\begin{array}{cc}
-e^{i\varphi}\left(e^{i\phi}\sin^{2}\theta\!+\!\cos^{2}\theta\right) & \left(1\!-\!e^{i\phi}\right)\sin\theta\cos\theta\\
e^{i\varphi}\left(1\!-\!e^{i\phi}\right)\sin\theta\cos\theta & -e^{i\phi}\cos^{2}\theta\!-\!\sin^{2}\theta
\end{array}\!\right]\!\!,\!\label{eq:Matrix_of_G}
\end{equation}
due to
\begin{eqnarray}
G\left|\alpha\right\rangle  & = & G_{11}\left|\alpha\right\rangle +G_{21}\left|\beta\right\rangle ,\\
G\left|\beta\right\rangle  & = & G_{12}\left|\alpha\right\rangle +G_{22}\left|\beta\right\rangle ,
\end{eqnarray}
where $G_{ij}$ refers to the entry in the $i$-th row and $j$-th
column of the matrix in Eq.~(\ref{eq:Matrix_of_G}).

Suppose the initial state is
\begin{equation}
\left|\psi\right\rangle
= H^{\otimes n}\left|0\right\rangle
= \sin\theta\left|\alpha\right\rangle
+\cos\theta\left|\beta\right\rangle,
\label{eq:initial_state}
\end{equation}
where $\theta=\arcsin\sqrt\lambda$, $\theta\in\left(0,\pi/2\right)$.
After $k$ iterations of $G\left(\phi,\varphi\right)$ with the phase
matching condition \cite{Long1999}
\begin{equation}
\phi=\varphi,\label{eq:phase_matching_of_Long1999}
\end{equation}
the state becomes \cite{Zhong2008}
\begin{equation}
G^{k}\left|\psi\right\rangle =a_{k}^{\phi}\left|\alpha\right\rangle +b_{k}^{\phi}\left|\beta\right\rangle ,
\end{equation}
where
\begin{equation}
a_{k}^{\phi}\!=\!\frac{\sin\theta}{\sin\delta}\left(\!-\!1\right)^{k}\!e^{i\left(k\!-\!1\right)\phi}
\!\left\{ e^{i\phi}\sin\!\left[\left(k\!+\!1\right)\delta\right]\!
-\!\sin\left(k\delta\right)\right\}\!,
\end{equation}
and
\begin{equation}\label{eq:expression_delta}
\delta=\arccos\left[1-\lambda\left(1-\cos\phi\right)\right]\in\left(0,\pi\right).
\end{equation}
The success probability of finding
the superposition of target states is thus given by
\begin{eqnarray}
P_{k}^{\phi}\left(\lambda\right) & = & \big|a_{k}^{\phi}\big|^{2}=A\cos\left[\left(2k+1\right)\delta\right]+B,\label{eq:success_rate_P_k_phi}
\end{eqnarray}
where
\begin{eqnarray}
A & = & \frac{\sin^{2}\theta}{\sin^{2}\delta}\left(\cos\phi-\cos\delta\right),\\
B & = & \frac{\sin^{2}\theta}{\sin^{2}\delta}\left(1-\cos\phi\cos\delta\right).
\end{eqnarray}
From Eq.~(\ref{eq:success_rate_P_k_phi}), we can see that $P_{k}^{\phi}\left(\lambda\right)=P_{k}^{2\pi-\phi}\left(\lambda\right)$,
and if $\phi=0$, then $G=-I$,
the initial state is just multiplied by a phase factor of $-1$.
Therefore, only $\phi\in\left(0,\pi\right]$ needs to be considered.

The condition that derivative of $P_{k}^{\phi}\left(\lambda\right)$
equal to zero gives rise to all the local maximum points of $P_{k}^{\phi}\left(\lambda\right)$
on the range of $0<\lambda<1$ (Proof see appendix~\ref{sub:all_local_maximum_points}),
\begin{equation}
\lambda_{k,j}^{\phi,max}=\frac{1-\cos\big(\frac{2j-1}{2k+1}\pi\big)}{1-\cos\phi},\thinspace 1\le j\le k.\label{eq:max_point_of_success_rate}
\end{equation}
In addition, we can find that $\lambda_{k,j}^{\phi,max}$ increases as $j$ grows, and $P_{k}^{\phi}\big(\lambda_{k,j}^{\phi,max}\big)=100\%$.
\section{Generalized complementary - multiphase search algorithm \label{sec:Generalized_complementary_multip}}

According to Eqs.~(\ref{eq:success_rate_P_k_phi}) and (\ref{eq:max_point_of_success_rate}),
it is found that the algorithm has advantage of high success probability
near its local maximum points, and has disadvantage of low success
probability near the corresponding local minimum points. Thus, it
is difficult to maintain high success probability over the entire
range of $\lambda$, by applying $k$ iterations with just a single
phase. One would naturally expect that this problem could be handled
by using multiple phases. The key idea of the complementary-multiphase
algorithm is that a phase is employed only in the $\lambda$ range
where the algorithm has high success probability. For a certain phase,
in the range where the algorithm has low success probability, we use
other phases to make up for it. In this way, we would expect that
complementing multiple phases with each other could improve the
overall minimum success probability of the algorithm
to be no less than any given $P_{cri}\in$$\left(0,1\right)$, similar to Ref.~\cite{Yoder2014}.
The model of algorithm is described in the following.

\subsection{Model \label{sub:Model}}

We first divide the entire range of $\lambda\in\left(0,1\right)$
into a series of small ranges, denoted by $\varLambda_{1}$, $\varLambda_{2}$,
$\cdots$, $\varLambda_{k}$, $\cdots$, satisfying the relation
\begin{equation}
\bigcup\limits _{k\ge1}\varLambda_{k}=\left(0,1\right).
\end{equation}
For each
$\varLambda_{k}$, we specify the number of iterations of the algorithm
to be $k$. Then by subdividing range $\varLambda_{k}$ further,
we get smaller ranges, denoted by $\varLambda_{k,1}$, $\varLambda_{k,2}$,
$\cdots$, $\varLambda_{k,n_{k}}$,
satisfying
\begin{equation}
\bigcup\limits _{m=1}^{n_{k}}\varLambda_{k,m}=\varLambda_{k}.
\end{equation}
In this way, the entire range $\left(0,1\right)$ is finally divided
into $\varLambda_{1,1}$, $\varLambda_{1,2}$, $\cdots$, $\varLambda_{1,n_{1}}$,
$\cdots$, $\varLambda_{k,1}$, $\varLambda_{k,2}$, $\cdots$,
$\varLambda_{k,n_{k}}$, $\cdots$, with
\begin{equation}
\bigcup\limits _{k\ge1}\bigcup\limits _{m=1}^{n_{k}}\varLambda_{k,m}=\left(0,1\right).
\end{equation}

For each $\varLambda_{k,m}$, we specify the phase of the algorithm
to be $\phi_{k,m}$, such that the algorithm has a high success probability
no less than the given $P_{cri}$,
where $\phi_{k,m}\in\left(0,\pi\right]$, $1\le m\le n_{k}$.
Assuming for now the existence of $\varLambda_{k}$, $\varLambda_{k,m}$,
$\phi_{k,m}$ and $n_{k}$ --- their values are given later --- then, the
specific steps of the complementary-multiphase algorithm can be described
as follows:

Step 1: The phase and number of iterations of the algorithm
can be specified. In Case-KIGR,
the range that $\lambda$ belongs to
can be determined from the given ranges
$\{\varLambda_{k,m}:k\ge1,1\le m\le n_{k}\}$,
without loss of generality, denoted by $\varLambda_{k,m}$. Then
we obtain that the corresponding
phase and number of iterations
is $\phi_{k,m}$ and $k$, respectively.
Note that, Case-KPV where $\lambda$ is known precisely, is a subcase of Case-KIGR,
as shown in Fig.~\ref{fig:relationships_among_Cases}.
Thus, in Case-KPV, $k$ and $\phi_{k,m}$ can be obtained in the same way.

Step 2: Prepare the initial state to be the equal superposition state,
i.e., $\left|\psi\right\rangle =H^{\otimes n}\left|0\right\rangle $.

Step 3: Repeat application $k$ times of the Grover iteration with
arbitrary phases $G\left(\phi,\varphi\right)$ to the initial state
$\left|\psi\right\rangle $, with the condition $\phi=\varphi=\phi_{k,m}$.

Step 4: Measure the final state $G^{k}\left|\psi\right\rangle $.
This will produce one of the marked states with high success probability.

\subsection{Optimal parameters \label{sub:Optimal_parameters}}

The selection method of optimal parameters $\varLambda_{k}$, $\varLambda_{k,m}$,
$\phi_{k,m}$ and $n_{k}$ are given in the following.

First, as illustrated in the model of algorithm, $k$ iterations corresponds
to the range $\varLambda_{k}$, and therefore, different choices
of $\{\varLambda_{k}:k\ge1\}$ result in different iterations of the algorithm.
We define the optimal $\{\varLambda_{k}\}$ as the one that
makes the number of iterations as few as possible and also
enables the success probability to be no less than any given $P_{cri}$.
Indeed, such optimal $\{\varLambda_{k}\}$ exists, which can
be written in the form (Proof see appendix~\ref{sub:optimal_varLambda_k}),
\begin{equation}
\varLambda_{k}=\left[\lambda_{k,1}^{\pi,max},\lambda_{k-1,1}^{\pi,max}\right),
k\ge 1,
\label{eq:optimal_varLambda_k}
\end{equation}
where, consistent with Eq.~(\ref{eq:max_point_of_success_rate}),
\begin{equation}
\lambda_{k,1}^{\pi,max}\equiv\begin{cases}
1, & {\rm if}\thinspace k=0,\\
\lambda_{k,j=1}^{\phi=\pi,max}=\sin^{2}\frac{\pi}{4k+2}, & {\rm if}\thinspace k\ge1.
\end{cases}\label{eq:max_point_piphi_1st}
\end{equation}
For any $\lambda\in\varLambda_{k}$, it can be found that the scope
of possibly used phases by the multiphase-complementing method can
be further reduced from $\left(0,\pi\right]$ to $(\phi_{k}^{min},\pi]$
(Proof see appendix~\ref{sub:possibly_used_phases}), where
\begin{equation}
\phi_{k}^{min}=\arccos\left(1-\frac{2-2\cos\frac{\pi}{2k+1}}{1-\cos\frac{\pi}{2k-1}}\right).\label{eq:phi_min_on_varLambda_k}
\end{equation}
And, we can see that
the success probability $P_{k}^{\phi}\left(\lambda\right)$,
for any $\phi\in(\phi_{k}^{min},\pi]$,
has the following extreme properties
on the range $\varLambda_{k}$
(Proof see appendix~\ref{sub:extreme_properties_of_P_k_phi}).
\begin{property}
(1) For $\phi\in(\phi_{k}^{min},\pi]$ and $\lambda\in\varLambda_{k}$, when $k\ge1$,   $P_{k}^{\phi}\left(\lambda\right)$ has one and only one local maximum
point, denoted by
\begin{equation}
\lambda_{k,1}^{\phi,max}=\frac{1-\cos\frac{\pi}{2k+1}}{1-\cos\phi}.
\label{eq:lambda_k_1_phi_max}
\end{equation}
(2) For $\phi\in(\phi_{k}^{min},\pi]$ and $\lambda\in\varLambda_{k}$, when
$k=1$, $P_{k}^{\phi}\left(\lambda\right)$ has one and only one local minimum
point, denoted by
\begin{equation}\label{eq:lambda_k_1_phi_min}
\lambda_{k=1,1}^{\phi,min}=\frac{5-4\cos\phi}{6-6\cos\phi}.
\end{equation}
While, when $k\ge2$, there are no local minimum points.
\label{pro:extreme_properties_of_P_k_phi}
\end{property}

Secondly, according to the model of algorithm,
multiple phases are employed on $\varLambda_{k}$
by the multiphase-complementing method.
Assuming for now the phases are already known
--- the optimal values are given later ---
we denote these phases in descending order by
$\phi_{k,1}$, $\phi_{k,2}$, $\cdots$, $\phi_{k,n_{k}}$,
where $\phi_{k,m}\in(\phi_{k}^{min},\pi]$, $1\le m\le n_k$,
$n_{k}$ is defined to be the number of phases used on $\varLambda_{k}$.
The phase $\phi_{k,m}$ corresponds to the range $\varLambda_{k,m}$, namely,
$\phi_{k,m}$ is always used by the algorithm for any $\lambda\in\varLambda_{k,m}$.
Therefore, different choices of $\{\varLambda_{k,m}:1 \le m \le n_k\}$
result in different success probabilities of the algorithm.
We define the optimal $\{\varLambda_{k,m}\}$ as the one
yielding the largest minimum success probability on $\varLambda_{k}$.
Actually, based on Property \ref{pro:extreme_properties_of_P_k_phi},
we can see that such optimal $\{\varLambda_{k,m}\}$ exists,
and is given in the following form
(Proof see appendix~\ref{sub:optimal_varLambda_k_m}),
\begin{equation}
\varLambda_{k,m}=\left[a_{k,m-1},a_{k,m}\right),
1 \le m \le n_k,
\label{eq:optimal_varLambda_k_m}
\end{equation}
where $a_{k,m}$ denotes the point of intersection of the curves represented
by $P_{k}^{\phi_{k,m}}\left(\lambda\right)$ and $P_{k}^{\phi_{k,m+1}}\left(\lambda\right)$
for $1\le m\le n_{k}-1$, $a_{k,m}\equiv\lambda_{k,1}^{\pi,max}$
for $m=0$, and $a_{k,m}\equiv\lambda_{k-1,1}^{\pi,max}$ for $m=n_{k}$.

Thirdly, as seen from Eq.~(\ref{eq:optimal_varLambda_k_m}),
the optimal $\{\varLambda_{k,m}\}$ depends on the multiple phases used on $\varLambda_{k}$.
Therefore, different choices of
$\{\phi_{k,m}:1\le m\le n_{k}\}$
result in different $\{\varLambda_{k,m}\}$ and further
different minimum success probabilities.
We define the optimal $\{\phi_{k,m}\}$
as the one yielding the largest minimum success probability on $\varLambda_{k}$.
It is easy to see that the exhaustive method
to search the optimal $\{\phi_{k,m}\}$ is computationally infeasible,
because the exhaustion scale of all the
$\phi_{k,m}\in\left(\phi_{k}^{min},\pi\right]$
is infinitely large.
Fortunately, based on Property \ref{pro:extreme_properties_of_P_k_phi},
we find out the sufficient and necessary condition of optimal phases,
as shown in Theorem \ref{th:optimal_phases_condition}
(Proof see appendix~\ref{sub:optimal_phases_condition}).

\begin{theorem}
For the range of $\lambda\in\varLambda_{k}$, $k\ge1$,
assuming that the number of phases $n_{k}$ is given,
then we can get the sufficient and necessary condition
of the optimal $\{\phi_{k,m}:1\le m\le n_{k}\}$
as follows:
\begin{eqnarray}
P_{k}^{\phi_{k,1}}\left(a_{k,0}\right) & = & \cdots=P_{k}^{\phi_{k,m}}\left(a_{k,m-1}\right)  =  \cdots \nonumber \\
&=&P_{k}^{\phi_{k,n_{k}}}\left(a_{k,n_{k}-1}\right)=P_{k}^{\phi_{k,n_{k}}}\left(\lambda_{key}\right),\label{eq:optimal_phases_conditions}
\end{eqnarray}
where,
\begin{equation}
\lambda_{key}\equiv\begin{cases}
\lambda_{k,1}^{\phi_{k,n_{k}},min} & {\rm if}\thinspace k = 1,\\
\lambda_{k-1,1}^{\pi,max} & {\rm if}\thinspace k \ge 2.
\end{cases}\label{eq:optimal_phases_conditions_lam_key}
\end{equation}
\label{th:optimal_phases_condition}
\end{theorem}

Lastly, Eq.~(\ref{eq:optimal_phases_conditions}) gives
a set of $n_{k}$ equations about $\phi_{k,1}$, $\cdots$, $\phi_{k,n_{k}}$,
therefore, different choices of $n_{k}$ result in
different optimal phases and eventually
different success probabilities on $\varLambda_{k}$.
Note that, the larger $n_k$, the more densely $\varLambda_k$ being divided,
which makes the identification of the range
that $\lambda$ belongs to from the given ranges $\{\varLambda_{k,m}\}$
become more difficult.
Therefore, we define the optimal $n_{k}$ as the least integer
that meets our expectation, i.e.,
the success probability for any $\lambda\in\left(0,1\right)$
could be no less than any given $P_{cri}\in\left(0,1\right)$.
In order to determine the optimal $n_{k}$,
we first define $Q_{k}^{\pi}$ as the largest minimum success probability
on range $\varLambda_{k}$,
and afterwards get the following property of $Q_{k}^{\pi}$
with respect to the number of phases $n_{k}$
(Proof see appendix \ref{sub:properties_of_Q_k_pi}).
\begin{property}
(1) $Q_{k}^{\pi}$ increases as $n_{k}$ grows.

\noindent
(2) $Q_{k}^{\pi}\to100\%$ when $n_{k}\to\infty$.
\label{pro:properties_of_Q_k_pi}
\end{property}
Based on Property \ref{pro:properties_of_Q_k_pi},
the optimal $n_{k}$ can be determined as follows.

Step 1: Initialize $n_{k}=1$.

Step 2: According to the value of $n_{k}$ and the optimal phases condition
Eq.~(\ref{eq:optimal_phases_conditions}), calculate the largest minimum
success probability on $\varLambda_{k}$, namely $Q_{k}^{\pi}\left(n_{k}\right)$.

Step 3: Check whether $Q_{k}^{\pi}\left(n_{k}\right)$ is smaller
than $P_{cri}$. If $Q_{k}^{\pi}\left(n_{k}\right)<P_{cri}$, then
increase $n_{k}$ by one, and go back to Step 2; otherwise, output
$n_{k}$ as the optimal number of phases and abort the procedure.

At this point, we have obtained all the selection methods of
the optimal $\varLambda_{k}$, $\varLambda_{k,m}$, $\phi_{k,m}$ and $n_{k}$.
For clarity, below we make the complete selection process explicit.

First, according to Eq.~(\ref{eq:optimal_varLambda_k}),
we have a division of the entire range of $\lambda\in\left(0,1\right)$,
i.e.,
\begin{eqnarray}
\varLambda_{1}&=&\bigg[\frac{1}{4},1\bigg), \varLambda_{2}=\bigg[\frac{3-\sqrt{5}}{8},\frac{1}{4}\bigg), \cdots, \nonumber \\
\varLambda_{k}&=&\bigg[\sin^{2}\frac{\pi}{4k+2},\sin^{2}\frac{\pi}{4k-2}\bigg),\cdots.
\end{eqnarray}
In Case-KIGR, one can identify which of the given ranges
$\{\varLambda_{k}: k\ge1\}$ that $\lambda$ belongs to.
Without loss of generality, denote it by $\varLambda_{k}$.

Then, we can determine that the number of iterations of the algorithm is $k$.
After that, the optimal number of phases on $\varLambda_{k}$, denoted by $n_{k}$,
can be obtained for the given $P_{cri}$ and the above $k$.
With $n_k$, solving Eq.~(\ref{eq:optimal_phases_conditions})
will give rise to the optimal phases on $\varLambda_{k}$,
denoted by $\{\phi_{k,m}:1\le m\le n_{k}\}$,
which further yields the optimal $\{\varLambda_{k,m}:1 \le m \le n_k\}$
through Eq.~(\ref{eq:optimal_varLambda_k_m}).
In Case-KIGR, for the given ranges $\left\{\varLambda_{k,m}\right\}$,
the range that $\lambda$ belongs to,
without loss of generality denoted by $\varLambda_{k,m}$, can be identified.
Correspondingly, the phase of the algorithm can be finally specified as $\phi_{k,m}$.

Executing the algorithm with optimal parameters leads directly
to results of the success probability and number of iterations,
as is described in the following section.

\section{Analysis of performance \label{sec:Analysis_of_performance}}

\subsection{Success probability \label{sub:Success_probability}}

For our complementary-multiphase algorithm, on the one hand, $\varLambda_{1}$,
$\varLambda_{2}$, $\cdots$, $\varLambda_{k}$, $\cdots$ constitute
a division of the entire range of $\lambda\in\left(0,1\right)$. On
the other hand, on each $\varLambda_{k}$, the algorithm uses multiple
phases to complement each other, and the largest minimum success probability
on $\varLambda_{k}$ converges to 100\% when the number of phases
increases to infinity. It follows that, the success probability of
our algorithm is possible to be no less than any given $P_{cri}\in\left(0,1\right)$
for the entire range of $\lambda$.

\begin{figure}[t]
\begin{centering}
\includegraphics[width=7.5cm]{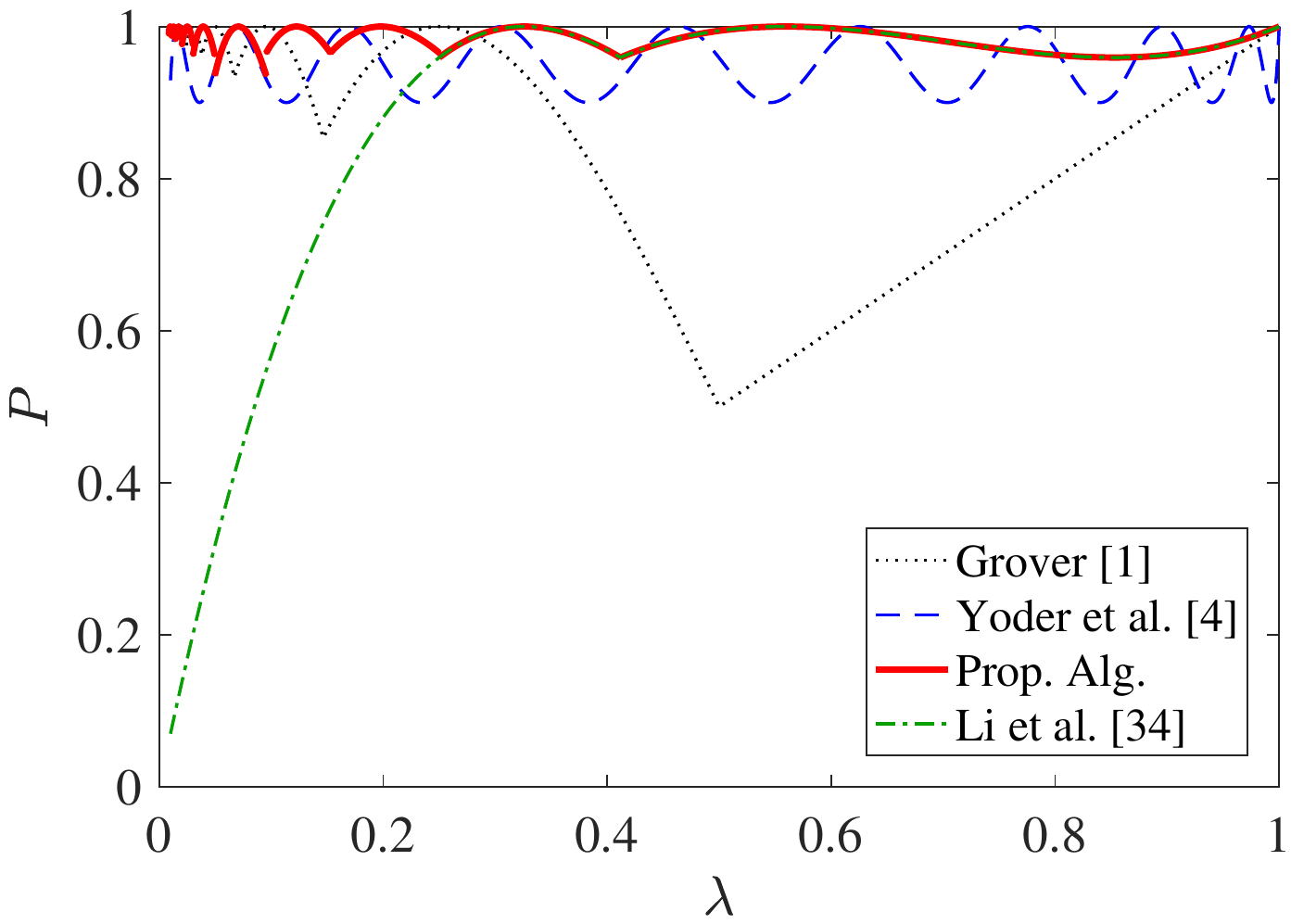}
\par\end{centering}
\protect\caption{The success probabilities $P$ as functions of
the fraction of target items $\lambda$.
The black dotted, blue dashed, red solid and green dashed-dotted curves
correspond to the original Grover algorithm \cite{Grover1996},
the optimal fixed-point algorithm \cite{Yoder2014},
our proposed algorithm and
the complementary-multiphase algorithm with only one iteration \cite{Li2014}, respectively.
The task is to achieve success probability no less than $P_{cri}=90\%$
for all $\lambda\ge\lambda_{0}=10^{-2}$.
\label{fig:success_probabilities_functions}}
\end{figure}

The success probabilities of the Grover algorithm \cite{Grover1996},
the optimal fixed-point algorithm \cite{Yoder2014},
our proposed algorithm, and
the complementary-multiphase algorithm with only one iteration \cite{Li2014}
as functions of $\lambda$
are presented in Fig.~\ref{fig:success_probabilities_functions},
with the fraction of target items $\lambda\ge\lambda_{0}=10^{-2}$
and the acceptable success probability $P\ge P_{cri}=90\%$.
As seen in Fig.~\ref{fig:success_probabilities_functions},
the problem of Grover's algorithm \cite{Grover1996}
that high success probability
over the entire range of $\lambda$ cannot be maintained
is systematically solved by our complementary-multiphase algorithm,
which overcomes the limitation of the applicable range in Ref.~\cite{Li2014}
where only $\lambda\in[\frac{1}{4},1)$ could be covered,
and achieves the same effect as the optimal fixed-point algorithm \cite{Yoder2014}.
By ``the same effect'', we mean
the success probability can be no less than
any given $P_{cri}\in\left(0,1\right)$ over the entire range.

The optimal parameters on each $\varLambda_{k}$ ($1\le k\le8$)
corresponding to Fig.~\ref{fig:success_probabilities_functions}
are given in Table~\ref{tab:optimal_parameters_in_Figure},
\begin{table}[b]
\begin{centering}
\protect\caption{The optimal multiple phases $\phi_{k,1},\cdots,\phi_{k,n_{k}}$ and
the largest minimum success probability $Q_{k}^{\pi}$ on $\varLambda_{k}$
in Fig.~\ref{fig:success_probabilities_functions}, for $1\le k\le8$.
\label{tab:optimal_parameters_in_Figure}}
\begin{tabular}{>{\centering}m{0.8cm}>{\centering}m{2.9cm}>{\centering}m{2.3cm}>{\centering}m{1.5cm}}
\hline
\hline
$k$ & $\varLambda_{k}$ & $\phi_{k,1},\cdots,\phi_{k,n_{k}}$ & $Q_{k}^{\pi}$   \bigstrut \tabularnewline
\hline
1 & $\left[0.25,1\right)$ & 2.134,1.465 & 0.9593\tabularnewline
2 & $\left[0.09549,0.25\right)$ & 2.163,1.536 & 0.9654\tabularnewline
3 & $\left[0.04952,0.09549\right)$ & 1.984 & 0.9354\tabularnewline
4 & $\left[0.03015,0.04952\right)$ & 2.137 & 0.9625\tabularnewline
5 & $\left[0.02025,0.03015\right)$ & 2.243 & 0.9757\tabularnewline
6 & $\left[0.01453,0.02025\right)$ & 2.322 & 0.9830 \tabularnewline
7 & $\left[0.01093,0.01453\right)$ & 2.383  & 0.9875 \tabularnewline
8 & $\left[0.008513,0.01093\right)$ & 2.432  & 0.9904\tabularnewline
\hline
\hline
\end{tabular}
\par\end{centering}
\end{table}
including:
the optimal multiple phases, denoted by $\phi_{k,1},\cdots,\phi_{k,n_{k}}$
and the largest minimum success probability, denoted by $Q_{k}^{\pi}$.
We can see that
the multiphase-complementing method indeed
guarantee a range of $\lambda\ge\lambda_0$ over which
the expectation $P\ge P_{cri}$ can be satisfied.

\subsection{Number of iterations \label{sub:Number_of_iterations}}

As described in the model of algorithm,
the number of iterations is specified by $k$ for any $\lambda\in\varLambda_{k}$,
and the optimal $\varLambda_{k}=\big[\lambda_{k,1}^{\pi,max},\lambda_{k-1,1}^{\pi,max}\big)$
is defined by Eq.~(\ref{eq:optimal_varLambda_k}).
Therefore, we have
\begin{eqnarray}
\lambda\in\varLambda_{k} & \Leftrightarrow & \lambda_{k,1}^{\pi,max}\le\lambda<\lambda_{k-1,1}^{\pi,max}\nonumber \\
 & \Leftrightarrow & k-\frac{1}{2}<\frac{\pi}{4\theta}\le k+\frac{1}{2},
\end{eqnarray}
where $\theta=\arcsin\sqrt{\lambda}$.
Thus, the number of iterations of our algorithm is given as
\begin{equation}
k=CI\left(\frac{\pi}{4\arcsin\sqrt{\lambda}}\right),\label{eq:optimal_k_of_us}
\end{equation}
where $CI\left(x\right)=k$ corresponds to $k-\frac{1}{2}<x\le k+\frac{1}{2}$.
From Eq.~(\ref{eq:optimal_k_of_us}), we also see that when $\lambda=M/N\ll1$, $k\approx\frac{\pi}{4}\sqrt{{N}/{M}}$,
due to $\arcsin\sqrt{\lambda}\approx\sqrt{\lambda}$.

\begin{figure}[tb]
\begin{centering}
\includegraphics[width=7.5cm]{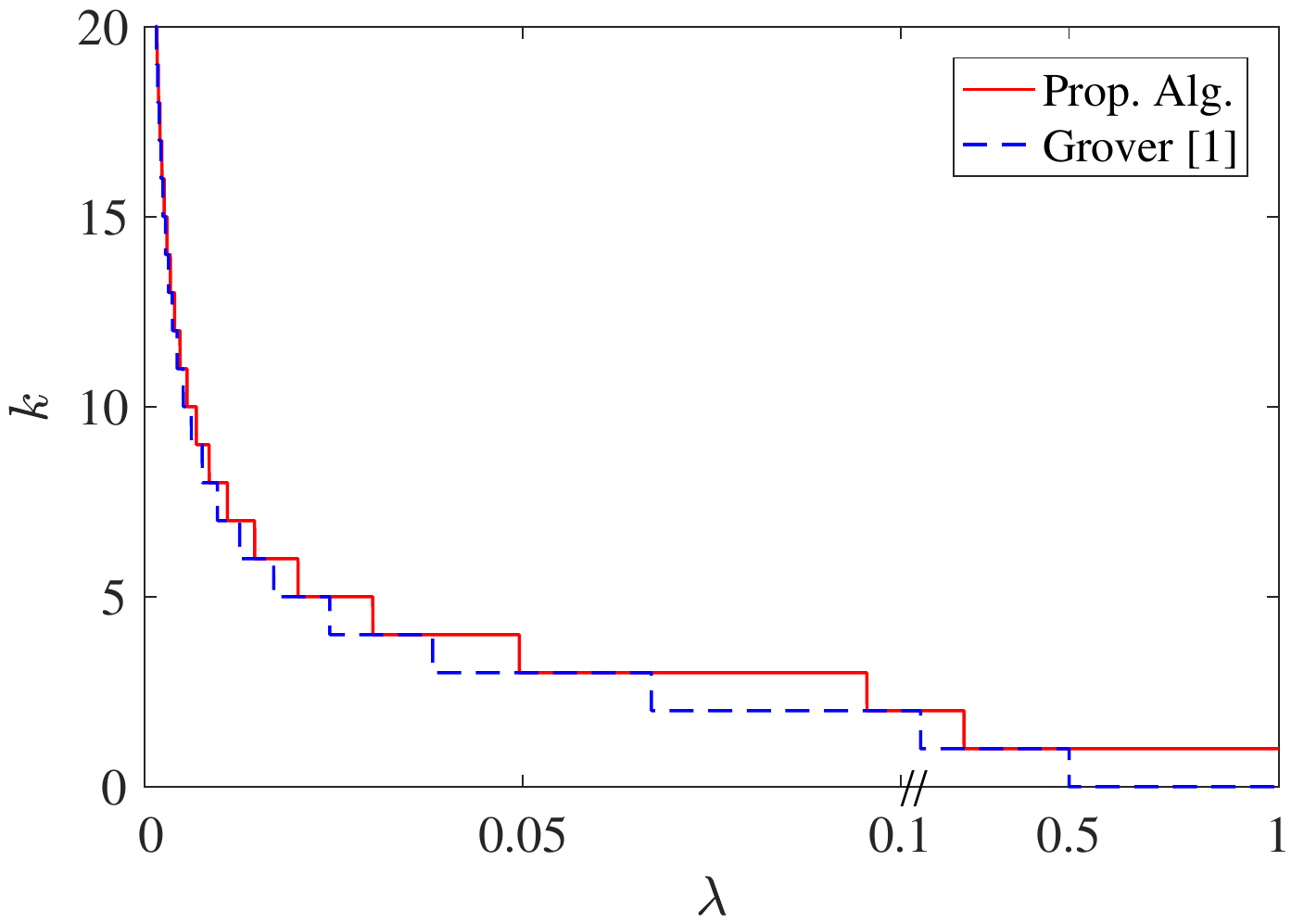}
\par\end{centering}
\protect\caption{{The number of iterations
as a function of the fraction of target items.
The red solid and blue dashed curves correspond to
our algorithm and the Grover algorithm \cite{Grover1996}, respectively.
Note that, to see the range of small $\lambda$ more clearly,
the range of large $\lambda$ is compressed,
with a ``$/\!/$'' on the x-axis marking the boundary.
\label{fig:number_of_iterations_function}}}
\end{figure}

From Eqs.~(\ref{eq:optimal_k_of_Grover_CI}) and (\ref{eq:optimal_k_of_us}),
it follows that,
\begin{equation}
k=\begin{cases}
k_{G}, & {\rm if}\thinspace\lambda\in\bigcup\limits _{k\ge1}\left[\sin^{2}\frac{\pi}{4k+2},\sin^{2}\frac{\pi}{4k}\right),\\
k_{G}+1, & {\rm if}\thinspace\lambda\in\bigcup\limits _{k\ge1}\left[\sin^{2}\frac{\pi}{4k},\sin^{2}\frac{\pi}{4k-2}\right).
\end{cases}\label{eq:comparison_of_k_kG}
\end{equation}
Figure~\ref{fig:number_of_iterations_function}
shows a comparison between the number of iterations of our algorithm $k$
and that of Grover's algorithm $k_{G}$ versus
the fraction of target items $\lambda$.
As can be seen, $k$ and $k_{G}$ are almost the same.

\section{Discussions \label{sec:Discussions}}

In this section, we will give some comparisons between
our algorithm and several other kinds of quantum search algorithms.

Compared with
the 100\%-success probability algorithms \cite{Brassard2002,Long2001,Toyama2013},
the fixed-phase algorithms \cite{Younes2013,Zhong2008,Li2012} and
the matched-multiphase algorithms \cite{Yoder2014,Toyama2008,Toyama2009},
the sequence of operations (denoted by $S$)
applied to the initial state
in our complementary-multiphase algorithm is significantly different.
The reasons are as follows:
Among the 100\%-success probability algorithms, Refs.~\cite{Long2001,Toyama2013}
repeat the same Grover iteration with arbitrary phases $k$ times,
with the sequence of operations being
\begin{equation}
S=G^{k}\left(\phi,\phi\right),\label{eq:operation_sequence_of_FixedP}
\end{equation}
where $k$ is first specified, satisfying
\begin{equation}
k\ge\left\lceil \frac{\pi}{4\arcsin\sqrt{\lambda}}-\frac{1}{2}\right\rceil ,\label{eq:k_of_Long2001}
\end{equation}
and then $\phi$ is a function of $k$ and $\lambda$,
\begin{equation}
\phi=2\arcsin\left(\sin\left(\frac{\pi}{4k+2}\right)/\sqrt{\lambda}\right).
\label{eq:phase_of_Long2001}
\end{equation}
While Ref.~\cite{Brassard2002} first performs the standard Grover iteration
$k$ times, and then run one more generalized Grover iteration with
arbitrary phases. The sequence of operations is given by
\begin{equation}
S=G\left(\phi,\varphi\right)G^{k}\left(\pi,\pi\right),\label{eq:operation_sequence_of_Bra2002}
\end{equation}
where $k=\left\lfloor \frac{\pi}{4\theta}-\frac{1}{2}\right\rfloor $,
$\phi$ and $\varphi$ satisfy the condition
\begin{equation}
\!\cot\!\left(\left(2k\!+\!1\right)\theta\right)\!=\!e^{i\varphi}\!\sin\!\left(2\theta\right)\left(i\cot\left(\phi/2\right)\!-\!\cos\left(2\theta\right)\right)^{\!-\!1}\!.\!\label{eq:phase_matching_of_Bra2002}
\end{equation}
The sequence of operations in the fixed-phase algorithms \cite{Younes2013,Zhong2008,Li2012}
is exactly in the same form as Eq.~(\ref{eq:operation_sequence_of_FixedP}).
While at this time, $\phi$ is first specified, and then $k$ is
a function of $\phi$ and $\lambda$ with the optimal value being
defined by Eq.~(\ref{eq:optimal_k_of_Li2012}).
Based on the multiphase-matching
method, Refs.~\cite{Yoder2014,Toyama2008,Toyama2009} utilize a set of multiple
phases $\phi_{j}$ and $\varphi_{j}$ ($1\le j\le k$) satisfying
the condition $\phi_{j}=\varphi_{k-j+1}$ globally over the sequence
of operations, as shown below:
\begin{equation}
S=G\left(\phi_{k},\varphi_{k}\right)G\left(\phi_{k-1},\varphi_{k-1}\right)\cdots G\left(\phi_{1},\varphi_{1}\right).\label{eq:operation_sequence_of_MPM}
\end{equation}
However, our complementary-multiphase algorithm divides the entire
range of $\lambda$ into a series of small ranges,
denoted by $\left\{\varLambda_{k,m}: k\ge1, 1\le m\le n_{k}\right\}$.
For each $\varLambda_{k,m}$, an individual
phase is specified correspondingly. Therefore, the sequence of operations
in our algorithm is indeed different from other algorithms, which
can be written as
\begin{equation}
S=G^{k}\left(\phi_{k,m},\phi_{k,m}\right),\thinspace{\rm for}\thinspace\lambda\in\varLambda_{k,m},\label{eq:operation_sequence_of_MPC}
\end{equation}
where the optimal $\{\varLambda_{k,m}\}$ is defined by Eq.~(\ref{eq:optimal_varLambda_k_m}).

Table~\ref{tab:detailed_comparison_of_algorithms}
\begin{table*}[tb]
\centering{}\protect\caption{The detailed comparisons between our algorithm and other algorithms. \label{tab:detailed_comparison_of_algorithms}}
\begin{centering}
\begin{tabular}{>{\raggedright}m{2.8cm}>{\raggedright}m{3.4cm}>{\raggedright}m{3cm}>{\raggedright}m{3.4cm}>{\raggedright}m{4.4cm}}
\hline
\hline
Algorithm & Applicable range of $\lambda$  & Success probability  & Number of iterations  & Phase(s)  \bigstrut\tabularnewline
\hline
Prop. Alg. & $\left(0,1\right)$ & $\ge P_{cri}$ & Eq.~(\ref{eq:optimal_k_of_us}) & $\phi_{k,m}$ for $\lambda\in\varLambda_{k,m}$\tabularnewline
Li et al. \cite{Li2014} & $\left[1/4,1\right)$ & $\ge P_{cri}$ & 1 & $\phi_{1,m}$ for $\lambda\in\varLambda_{1,m}$\tabularnewline
Yoder et al.\cite{Yoder2014}  & $\left(0,1\right)$ & $\ge1-\delta^{2}$ & Eq.~(\ref{eq:optimal_k_of_yoder2014}) & $\phi_{1},\cdots,\phi_{k}$ for $\lambda\in\left(0,1\right)$\tabularnewline
Toyama et al.\cite{Toyama2008}  & $\left[0.1,1\right)$ & $\ge99.8\%$ & 6 & $\phi_{1},\cdots,\phi_{6}$ for $\lambda\in\left[0.1,1\right)$\tabularnewline
Toyama et al.\cite{Toyama2009}  & $\left[0.006,0.11\right]$  & $\ge99.2\%$  & 20 & $\phi_{1},\cdots,\phi_{20}$ for $\lambda\in$$\left[0.006,0.11\right]$ \tabularnewline
Grover \cite{Grover1996}  & $\left(0,1\right)$ & $\ge50\%$ & $k_G$, Eq.~(\ref{eq:optimal_k_of_Grover_CI}) & $\pi$ for $\lambda\in\left(0,1\right)$\tabularnewline
Younes \cite{Younes2013}  & $\left(0,1\right)$ & $\ge99.58\%$ & $\lfloor \phi/\sqrt{\lambda} \rfloor $ & $1.91684\pi$ for $\lambda\in\left(0,1\right)$\tabularnewline
Zhong et al.\cite{Zhong2008}  & $\left(0,1\right)$ & $\ge99.43\%$ & $\lfloor \frac{\pi}{2}/\sqrt{\lambda} \rfloor $ & $1.018$ for $\lambda\in\left(0,1\right)$\tabularnewline
Li et al. \cite{Li2012} & $\left(0,1\right)$ & $\ge99.38\%$  & Eq.~(\ref{eq:optimal_k_of_Li2012}) & $0.1\pi$ for $\lambda\in\left(0,1\right)$\tabularnewline
Long \cite{Long2001} & $\left(0,1\right)$ & $=100\%$ & Eq.~(\ref{eq:k_of_Long2001}) & Eq.~(\ref{eq:phase_of_Long2001})\tabularnewline
Boyer et al. \cite{Boyer1998} & $\left(0,3/4\right]$ & $=100\%$ & $\le 11 k_G$ in expected & $\pi$ for $\lambda\in\left(0,1\right)$ \tabularnewline
\hline
\hline
\end{tabular}
\par\end{centering}
\end{table*}
lists the performances of our algorithm and other algorithms
in respect of the applicable range of $\lambda$,
the success probability $P$, the number of iterations $k$ and the phase(s) $\phi$.
The main advantages of our algorithm over other algorithms
are discussed in detail as follows.

Firstly, in respect of the applicable range of $\lambda$,
our algorithm applies to the entire range $\left(0,1\right)$, which
is essentially the same as Refs.~\cite{Grover1996,Yoder2014,Younes2013,Zhong2008,Li2012,Long2001}
and broader than Refs.~\cite{Boyer1998,Toyama2008,Toyama2009,Li2014}. Especially
compared to Ref.~\cite{Li2014} which is only applicable to $\left[1/4,1\right)$,
the limitation there is overcome completely by considering a general number of iterations in our algorithm.

Secondly, in respect of the success probability $P$, as shown in Fig.~\ref{fig:success_probabilities_functions},
our algorithm achieves the same effect as Refs.~\cite{Yoder2014,Li2014} allowing
$P\ge P_{cri}\in\left(0,1\right)$, which is more flexible than Refs.~\cite{Grover1996,Younes2013,Zhong2008,Li2012,Toyama2008,Toyama2009}.
Moreover, as illustrated in Property \ref{pro:properties_of_Q_k_pi}, on $\varLambda_{k}$ the largest
minimum success probability $Q_{k}^{\pi}\to100\%$ when $n_{k}\to\infty$.
Thus, it is possible to asymptotically achieve the effect of certainty in Ref.~\cite{Long2001},
by the multiphase-complementing method.

Thirdly, in respect of the number of iterations $k$,
as depicted in Fig.~\ref{fig:number_of_iterations_function},
our algorithm performs almost the same iterations as
the original Grover algorithm \cite{Grover1996} with up to once more,
and therefore has fewer iterations than
the trial-and-error algorithm \cite{Boyer1998}
and the fixed-phase algorithms \cite{Younes2013,Zhong2008,Li2012} with $\phi\neq\pi$,
due to $\arcsin^{-1}\big(\sqrt{\lambda}\sin\frac{\phi}{2}\big)
>\arcsin^{-1}\sqrt{\lambda}$.
Moreover, it follows from
Eqs.~(\ref{eq:optimal_k_of_yoder2014}) and (\ref{eq:optimal_k_of_us})
that in problems where
the acceptable minimum success probability $P_{cri}$ is greater than 82.71\%,
our algorithm uses fewer number of iterations than
the optimal fixed-point algorithm \cite{Yoder2014},
because when $P_{cri}>82.71\%$,
\begin{equation}
\frac{\log\left(2/\delta\right)}{2\sqrt{\lambda}} >
\frac{\pi}{4\sqrt{\lambda}},
\end{equation}
where $\delta=\sqrt{1-P_{cri}}$.
For example, when $P_{cri}=99.25\%$,
the number of iterations of our algorithm
is just one half of that of Ref.~\cite{Yoder2014}.

Finally, in respect of the phases, when $\lambda\in\varLambda_{k,m}$, we can
always find a target state with high success probability no less than
$P_{cri}$ without tuning the phase, similar to Refs.~\cite{Toyama2008,Toyama2009}.
Moreover, our complementary-multiphase algorithm is applicable
to Case-KIGR even without the precise knowledge of $\lambda$, where the 100\%-success probability algorithms \cite{Brassard2002,Chi1999,Long2001,Toyama2013,Liu2014}
cannot work, indicating that our algorithm has a wider scope of applications.

To sum up, in Case-KIGR, our algorithm
systematically solves the problem in success probability of the Grover algorithm and also
preserves its advantages in the applicable range of $\lambda$
and number of iterations.

\section{Conclusion \label{sec:Conclusion}}

In summary, we have presented a complementary-multiphase quantum search algorithm
with general iterations,
to solve the success probability problem of the Grover algorithm in the case
(denoted by Case-KIGR), where
one can identify the range that $\lambda$ belongs to
from a given series of disjoint ranges of $\lambda$.
To improve the overall minimum success probability by complementing multiple phases,
we divided the entire range of $0<\lambda<1$ into a series of small ranges.
For each range, the number of iterations and phase of the algorithm
were individually specified.
Moreover, we derived all local maximum points
of the success probability after applying the Grover iteration
with arbitrary phases $k$ times, and further obtained the optimal division
of range $\left(0,1\right)$, denoted by $\left\{\varLambda_{k}:k\ge 1\right\}$,
that minimizes the query complexity \cite{Yoder2014} of quantum searching.
In addition, the extreme properties
of the success probability on range $\varLambda_{k}$ were analyzed,
and the optimal division of $\varLambda_{k}$, optimal number of phases, and optimal phases condition were subsequently obtained,
which maximize the minimum success probability of the algorithm.

Compared with the existing algorithms, in Case-KIGR, our algorithm simultaneously achieves
the following three goals for the first time:
(1) the success probability can be no less than any $P_{cri}\in\left(0,1\right)$,
(2) the entire range of $0<\lambda<1$ can be covered, and
(3) the required number of iterations can be almost the same as
the original Grover algorithm.
Especially when the required minimum success probability is no less than 82.71\%,
our algorithm uses fewer iterations than the optimal fixed-point algorithm \cite{Yoder2014}.
The multiphase-complementing method
provides a new idea for the research on quantum search algorithms.
Further investgation may be extended to the general case where one knows that $0 < \lambda < 1$.


\section{Acknowledgments \label{sec:Acknowledgments}}

We thank Ru-Shi He and Zheng-Mao Xu for useful discussions. This work
was supported by the National Natural Science Foundation of China
(Grant Nos. 11504430 and 61502526), and the National Basic Research
Program of China (Grant No. 2013CB338002).

\appendix \label{sec:Appendix}

\section{Proof of all local maximum points of $P_{k}^{\phi}\left(\lambda\right)$
on $\left(0,1\right)$ of Eq.~(\ref{eq:max_point_of_success_rate}) \label{sub:all_local_maximum_points}}

According to Eq.~(\ref{eq:success_rate_P_k_phi}), the derivative of
$P_{k}^{\phi}\left(\lambda\right)$ with respect to $\lambda$ can
be written as
\begin{equation}
\frac{\partial P_{k}^{\phi}}{\partial\lambda}=\left(1+\cos\delta\right)^{-2}\left(\!\frac{\partial P_{k}^{\phi}}{\partial\lambda}\!\right)^{\!\left(1\right)},\label{eq:first_derivative_of_success_rate}
\end{equation}
where
\begin{eqnarray}
\left(\frac{\partial P_{k}^{\phi}}{\partial\lambda}\right)^{\left(1\right)} & \!\!\!\equiv & \left(1+\cos\phi\right)\left\{ 1+\cos\left[\left(2k+1\right)\delta\right]\right\} \!-\!\left(2k+1\right)\nonumber \\
 &  & \times\left(\cos\phi-\cos\delta\right)\left(1+\cos\delta\right)
 \frac{\sin\left[\left(2k+1\right)\delta\right]}{\sin\delta}.\nonumber
\end{eqnarray}
If $\cos\left[\left(2k+1\right)\delta\right]=-1$, then $\sin\left[\left(2k+1\right)\delta\right]=0$,
$\frac{\partial P_{k}^{\phi}}{\partial\lambda}=0$ and $P_{k}^{\phi}=1$,
thus solving the equation $\cos\left[\left(2k+1\right)\delta\right]=-1$
gives rise to the local maximum points of $P_{k}^{\phi}\left(\lambda\right)$.
The corresponding solutions are given as $\lambda_{k,j}^{\phi,max}$
for $1\le j\le k$ in Eq.~(\ref{eq:max_point_of_success_rate}).

We have established the existence of local maximum points, and now
we can further show that there are no other points except for $\lambda_{k,j}^{\phi,max}$.
From De Moivre's theorem (See p.~9 of Ref.~\cite{Zwillinger2011}), i.e.,
\begin{equation}
\!\!\cos\!\left[\left(2k\!+\!1\right)\delta\right]+i\sin\!\left[\left(2k\!+\!1\right)\delta\right]\!=
\!\left(\cos\delta\!+\!i\sin\delta\right)^{2k\!+\!1}\!,\label{eq:demoivre_theorem}
\end{equation}
where $i=\sqrt{-1}$, it follows that with respect to $\cos\delta$, $\cos\left[\left(2k+1\right)\delta\right]$
and ${\sin\left[\left(2k+1\right)\delta\right]}/{\sin\delta}$ are
polynomials of degree $2k+1$ and $2k$, respectively.
Consequently, the degree of the polynomial $\big(\frac{\partial P_{k}^{\phi}}{\partial\lambda}\big)^{\!\left(1\right)}$
is no more than $2k+2$, which will have up to $2k+1$ real roots for $\delta\in\left(0,\pi\right)$, now that $\delta=\pi$ is already one of its roots.
Furthermore, due to $P_{k}^{\phi}\left(\lambda=0\right)=0$ and $P_{k}^{\phi}\left(\lambda=1\right)=1$,
$P_{k}^{\phi}\left(\lambda\right)$ has the same number of local maximum
points and local minimum points.
Finally, we are now in a position
to conclude that $\lambda_{k,j}^{\phi,max}$ ($1\le j\le k$) are just
all the local maximum points of $P_{k}^{\phi}(\lambda)$. $\hfill\blacksquare$

\section{Proof of the optimal $\varLambda_{k}$ of Eq.~(\ref{eq:optimal_varLambda_k})
\label{sub:optimal_varLambda_k}}

On the one hand, to ensure the success probability of the complementary-multiphase algorithm
can be no less than any given $P_{cri}$,
for any $\lambda\in\varLambda_{k}$ there should be a phase
such that after $k$ iterations 100\% success probability can be reached.
On the other hand, to make the number of iterations as few as possible,
there should be no such a phase with $k-1$ iterations.

From Eq.~(\ref{eq:max_point_of_success_rate}), it follows that
for any $j\ge1$, $\lambda_{k,j}^{\phi,max}\ge\lambda_{k,1}^{\phi,max}$,
and for any $\phi\in\left(0,\pi\right]$, $\lambda_{k,1}^{\phi,max}\ge\lambda_{k,1}^{\pi,max}$.
Then, for any $\lambda\in[\lambda_{k,1}^{\pi,max},1)$
(or $[\lambda_{k-1,1}^{\pi,max},1)$),
there exists a phase such that
the success probability reaches 100\% with $k$ (or $k-1$) iterations.
Therefore, the corresponding optimal range of $\lambda$ to $k$ iterations
can be given as
$\varLambda_{k}=[\lambda_{k,1}^{\pi,max},\lambda_{k-1,1}^{\pi,max})$ ($k\ge 1$),
which constitute a division of the entire range of $\lambda$. $\hfill\blacksquare$

\section{Proof of the scope of possibly used phases on $\varLambda_{k}$
\label{sub:possibly_used_phases}}

Based on Eq.~(\ref{eq:max_point_of_success_rate}), it is found that for any $\phi\in\left(0,\phi_{k}^{\min}\right]$,
\begin{equation}
\lambda_{k,1}^{\phi,max}\ge\lambda_{k-1,1}^{\pi,max},
\end{equation}
and for any $\phi\in\left(\phi_{k}^{\min},\pi\right]$,
\begin{equation}
\lambda_{k,1}^{\pi,max}\le\lambda_{k,1}^{\phi,max}<\lambda_{k-1,1}^{\pi,max},
\end{equation}
where $\phi_{k}^{min}$
is defined by Eq.~(\ref{eq:phi_min_on_varLambda_k}). Then, for any
$\lambda\in\varLambda_{k}$ and any $\phi\in\left(0,\phi_{k}^{\min}\right]$,
we have
\begin{equation}
P_{k}^{\phi}\left(\lambda\right)\le P_{k}^{\phi_{k}^{min}}\left(\lambda\right).
\end{equation}
Consequently, the possibly used phases on $\varLambda_{k}$ of the
multiphase-complementing method can be limited to $\left(\phi_{k}^{min},\pi\right]$. $\hfill\blacksquare$

\section{Proof of the extreme properties of $P_{k}^{\phi}$ on $\varLambda_{k}$ of Property \ref{pro:extreme_properties_of_P_k_phi}
\label{sub:extreme_properties_of_P_k_phi}}

(1) On one hand, as mentioned in Appendix \ref{sub:possibly_used_phases},
for any $\phi\in\left(\phi_{k}^{\min},\pi\right]$, we have $\lambda_{k,1}^{\phi,max}\in\varLambda_{k}=\big[\lambda_{k,1}^{\pi,max},\lambda_{k-1,1}^{\pi,max}\big)$.
On the other hand, it can be found that $\lambda_{k,j}^{\phi,max}\notin\varLambda_{k}$
for $j\ge2$. This is because $k\ge1$ and $j\ge2$ lead to $\cos\frac{3\pi}{2k+1}\le\cos\frac{\pi}{2k-1}$,
and then
\begin{eqnarray}
\lambda_{k,j}^{\phi,max}&\ge&\lambda_{k,2}^{\phi,max}
=\frac{1-\cos\frac{3\pi}{2k+1}}{1-\cos\phi}\nonumber \\
&\ge&\frac{1-\cos\frac{\pi}{2k-1}}{2}
=\lambda_{k-1,1}^{\pi,max}.
\end{eqnarray}
Therefore, $\lambda_{k,1}^{\phi,max}$ is the one and only one local
maximum point of $P_{k}^{\phi}$ on $\varLambda_{k}$.

(2) In the case of $k=1$, we obtain $\varLambda_{k}=\left[{1}/{4},1\right)$.
Then, from Eq.~(2.13) in Ref.~\cite{Toyama2008} or Eq.~(6) in Ref.~\cite{Li2014},
it is straightforward to show that $\lambda_{k,1}^{\phi,min}$ given in Eq.~(\ref{eq:lambda_k_1_phi_min})
is the one and only one local minimum point of $P_{k}^{\phi}$ on
$\varLambda_{k}$.

In the case of $k\ge2$, to prove $\lambda_{k,1}^{\phi,min}\ge\lambda_{k-1,1}^{\pi,max}$,
we only need to find a $\lambda_{mid}$ such that $\lambda_{k,1}^{\phi,min}\ge\lambda_{mid}$
and $\lambda_{mid}\ge\lambda_{k-1,1}^{\pi,max}$. Indeed, such $\lambda_{mid}$
exists and may be given in the form
\begin{equation}
\lambda_{mid}=\frac{2\sin^{2}\frac{\pi}{4k-2}}{1-\cos\phi}.
\end{equation}
Since $1-\cos\phi\le2$, we have
\begin{equation}
\lambda_{mid}\ge\sin^{2}\frac{\pi}{4k-2}=\lambda_{k-1,1}^{\pi,max}.
\end{equation}
It remains to show that $\lambda_{k,1}^{\phi,min}\ge\lambda_{mid}$,
which is equivalent to prove $\frac{\partial P_{k}^{\phi}}{\partial\lambda}\big|_{\lambda=\lambda_{mid}}\le0$,
due to for $k\ge2$,
\begin{equation}
\lambda_{k,1}^{\phi,max}\le\lambda_{mid}<\frac{2\sin^{2}\frac{3\pi}{4k+2}}{1-\cos\phi}=\lambda_{k,2}^{\phi,max}.
\end{equation}
Here we denote $\frac{\partial P_{k}^{\phi}}{\partial\lambda}\big|_{\lambda=\lambda_{mid}}$
to be the value of $\frac{\partial P_{k}^{\phi}}{\partial\lambda}$
at $\lambda_{mid}$. The proof is carried out as follows.

First, for $\lambda=\lambda_{mid}$, it follows from Eq.~(\ref{eq:expression_delta})
that $\delta=\frac{\pi}{2k-1}$, and $\delta\le\frac{\pi}{3}$ now
that $k\ge2$. Substituting $\delta$ into Eq.~(\ref{eq:first_derivative_of_success_rate}),
we get
\begin{equation}
\frac{\partial P_{k}^{\phi}}{\partial\lambda}\Big|_{\lambda=\lambda_{mid}}=\frac{2}{1+\cos\delta}\left(\!\frac{\partial P_{k}^{\phi}}{\partial\lambda}\!\right)^{\!\left(2\right)}\!,\label{eq:first_derivative_of_success_rate_lam_mid}
\end{equation}
where
\begin{eqnarray}
\!\left(\!\frac{\partial P_{k}^{\phi}}{\partial\lambda}\!\right)^{\!\left(2\right)}
 \!\!\equiv  \left(1\!+\!2k\cos\delta\right)\!\cos\phi
\!-\!\!\left[\left(2k\!+\!1\right)\!\cos^{2}\!\delta\!+\!\cos\delta\!-\!1\right]\!,\nonumber
\label{eq:first_derivative_of_success_rate_2}
\end{eqnarray}
from which we obtain $\big(\frac{\partial P_{k}^{\phi}}{\partial\lambda}\big)^{\!\left(2\right)}$is
a monotonically decreasing function with respect to $\phi$ for any
given $k$, yielding
\begin{equation}
\left(\!\frac{\partial P_{k}^{\phi}}{\partial\lambda}\!\right)^{\!\left(2\right)}<
\left(\!\frac{\partial P_{k}^{\phi}}{\partial\lambda}\!\right)^{\!\left(2\right)}
\Big|_{\phi=\phi_{k}^{min}},\label{eq:first_derivative_of_success_rate_2_le_phi}
\end{equation}
for $\phi\in\left(\phi_{k}^{min},\pi\right]$.

Next, according to Eq.~(\ref{eq:phi_min_on_varLambda_k}), $\big(\frac{\partial P_{k}^{\phi}}{\partial\lambda}\big)^{\!\left(2\right)}|_{\phi=\phi_{k}^{min}}$
is an univariate function of $k$. When $k$ is sufficiently large,
namely $k\to\infty$, $\cos\delta=\cos\frac{\pi}{2k-1}\approx1$ and
therefore,
\begin{equation}
\left(\!\frac{\partial P_{k}^{\phi}}{\partial\lambda}\!\right)^{\!\left(2\right)}
\Big|_{\phi=\phi_{k}^{min}}
\!\approx\left(1+2k\right)\left(\cos\phi_{k}^{min}-1\right)\!<\!0,
\label{eq:first_derivative_of_success_rate_2_phi}
\end{equation}
which can also be numerically proven to hold for small $k$, for example
$k=2,3,\cdots,1000$.

Finally, based on Eqs.~(\ref{eq:first_derivative_of_success_rate_lam_mid}), (\ref{eq:first_derivative_of_success_rate_2_le_phi}) and (\ref{eq:first_derivative_of_success_rate_2_phi}),
for $k\ge2$, we have
$\frac{\partial P_{k}^{\phi}}{\partial\lambda}\big|_{\lambda=\lambda_{mid}}<0$,
and therefore $\lambda_{k,1}^{\phi,min}>\lambda_{mid}\ge\lambda_{k-1,1}^{\pi,max}$, namely there exists no local minimum points on $\varLambda_{k}$
for $P_{k}^{\phi}$. $\hfill\blacksquare$

\section{Proof of the optimal $\varLambda_{k,m}$ of Eq.~(\ref{eq:optimal_varLambda_k_m})
\label{sub:optimal_varLambda_k_m}}

Based on the Property \ref{pro:extreme_properties_of_P_k_phi}, we can obtain
\begin{equation}
\lambda_{k,1}^{\phi_{k,1},max}<\lambda_{k,1}^{\phi_{k,2},max}<
\cdots<\lambda_{k,1}^{\phi_{k,n_{k}},max},
\end{equation}
due to the assumption of
\begin{equation}
\phi_{k,1}>\phi_{k,2}>\cdots>\phi_{k,n_{k}}.\nonumber
\end{equation}
Then,
on $\big[\lambda_{k,1}^{\phi_{k,m},max},\lambda_{k,1}^{\phi_{k,m+1},max}\big)$
for $1\le m\le n_{k}-1$,
$P_{k}^{\phi_{k,m}}\left(\lambda\right)$
monotonically decreases
while $P_{k}^{\phi_{k,m+1}}\left(\lambda\right)$
monotonically increases
and
$P_{k}^{\phi_{k,m}}\big(\lambda_{k,1}^{\phi_{k,m},max}\big)=100\%$,
$P_{k}^{\phi_{k,m+1}}\big(\lambda_{k,1}^{\phi_{k,m+1},max}\big)=100\%$.
According to the intermediate value theorem (See p.~271 of Ref.~\cite{Zwillinger2011}),
there exists a $\lambda\in\big(\lambda_{k,1}^{\phi_{k,m},max},\lambda_{k,1}^{\phi_{k,m+1},max}\big)$
such that
\begin{equation}
P_{k}^{\phi_{k,m}}\left(\lambda\right)=P_{k}^{\phi_{k,m+1}}\left(\lambda\right).
\label{eq:intersection_point_a_km}
\end{equation}
We denote the solution as $a_{k,m}$,
which represents the intersection point
of $P_{k}^{\phi_{k,m}}\left(\lambda\right)$ and
$P_{k}^{\phi_{k,m+1}}\left(\lambda\right)$
on $\varLambda_{k}$.

Consequently, to maximize the minimum success probability of the algorithm
by taking advantage of the multiple phases,
$\phi_{k,1}$, $\phi_{k,m}$, $\phi_{k,m+1}$,  and $\phi_{k,n_{k}}$
should be employed on
$\big[\lambda_{k,1}^{\pi,max}\equiv a_{k,0},\lambda_{k,1}^{\phi_{k,1},max}\big)$,
$\big[\lambda_{k,1}^{\phi_{k,m},max},a_{k,m}\big)$,
$\big[a_{k,m},\lambda_{k,1}^{\phi_{k,m+1},max}\big)$
and $\big[\lambda_{k,1}^{\phi_{k,n_{k}},max},
\lambda_{k-1,1}^{\pi,max}\equiv a_{k,n_{k}}\big)$
respectively, where $1\le m\le n_{k}-1$.
Finally, the range of $\lambda$ corresponding to
$\phi_{k,m}$ ($1\le m\le n_{k}$) can be written as
\begin{eqnarray}
\varLambda_{k,m}&=&\left[a_{k,m-1},\lambda_{k,1}^{\phi_{k,m},max}\right)
\bigcup\left[\lambda_{k,1}^{\phi_{k,m},max},a_{k,m}\right)\nonumber \\
&=&\Big[a_{k,m-1},a_{k,m}\Big),
\end{eqnarray}
as desired. $\hfill\blacksquare$

\section{Proof of the optimal phases condition of Eq.~(\ref{eq:optimal_phases_conditions})
\label{sub:optimal_phases_condition}}

\begin{figure}[tb]
\begin{centering}
\includegraphics[width=7.5cm]{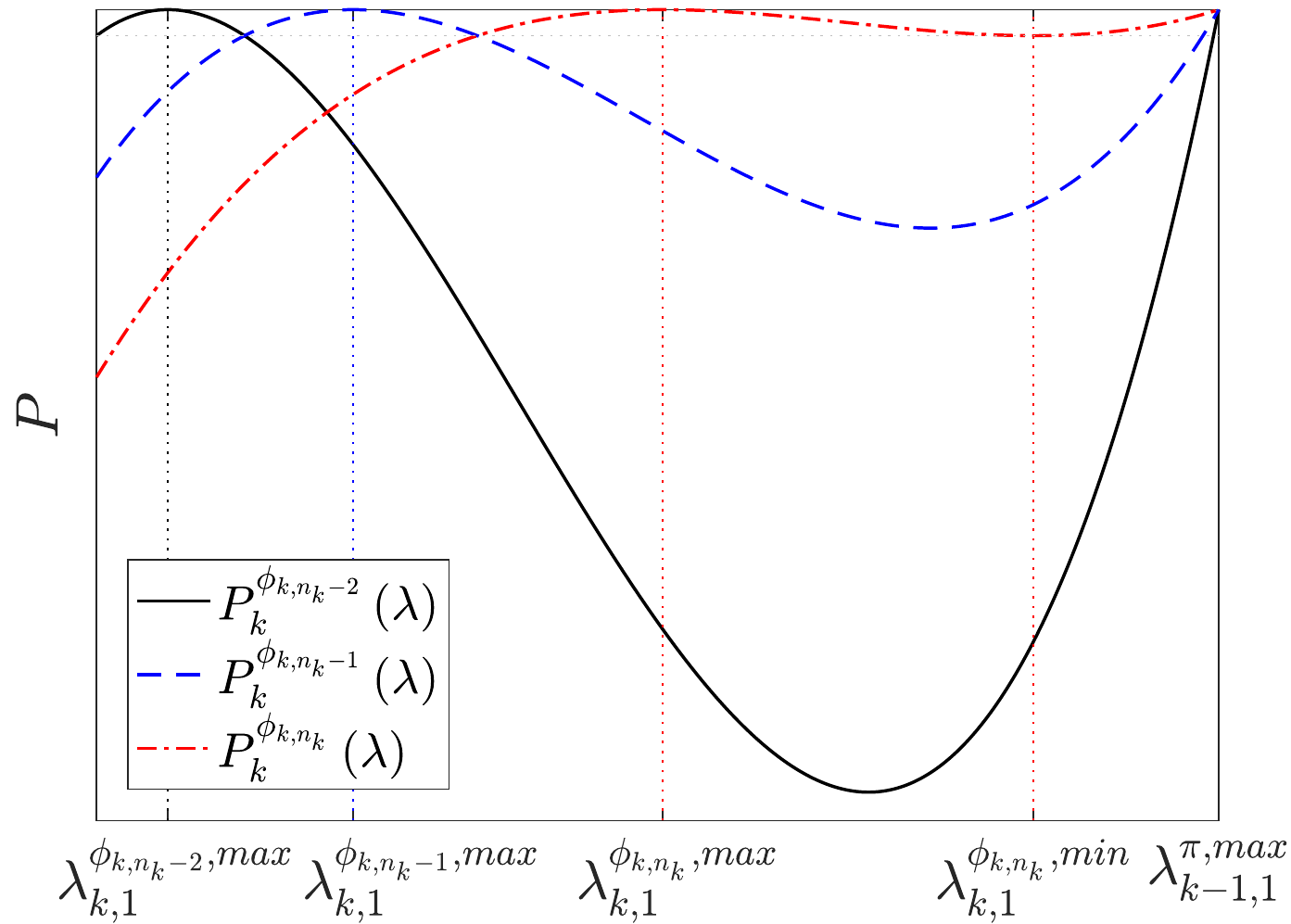}
\par\end{centering}

\protect\caption{The schematic of success probability $P$
as a function of the fraction of target items $\lambda$.
The black solid, blue dashed,
red dashed-dotted curves correspond to $\phi_{k,n_{k}-2}$, $\phi_{k,n_{k}-1}$ and $\phi_{k,n_{k}}$, respectively,
where
$\lambda_{k,1}^{\phi,max}$ defined by Eq.~(\ref{eq:lambda_k_1_phi_max}), and
$\lambda_{k=1,1}^{\phi,min}$ defined by Eq.~(\ref{eq:lambda_k_1_phi_min}) represent the
the local maximum point and minimum point of $P^{\phi}_k\left(\lambda\right)$, respectively.
\label{fig:schematic_for_optimal_phases}}
\end{figure}

In the case of $k=1$, first we can show that for any $\phi_{k,n_{k}-1}$, the optimal phases condition to maximize the minimum success probability of the algorithm on $\big[\lambda_{k,1}^{\phi_{k,n_{k}-1},max},\lambda_{k-1,1}^{\pi,max}\big)$ is
\begin{equation}
P_{k}^{\phi_{k,n_{k}}}\big(a_{k,n_{k}-1}\big)=P_{k}^{\phi_{k,n_{k}}}\big(\lambda_{k,1}^{\phi_{k,n_{k}},min}\big).
\label{eq:proof_opt_phases_keq1_phi_nk_1}
\end{equation}
Note that, $\lambda_{k,1}^{\phi,max}$ and $\lambda_{k=1,1}^{\phi,min}$ are defined by Eqs.~(\ref{eq:lambda_k_1_phi_max}) and (\ref{eq:lambda_k_1_phi_min}) respectively, and $a_{k,m}$ is the solution of Eq.~(\ref{eq:intersection_point_a_km}).
This is because, for a given $\phi_{k,n_{k}-1}$,
the minimum success probability on $\big[\lambda_{k,1}^{\phi_{k,n_{k}-1},max},\lambda_{k-1,1}^{\pi,max}\big)$
is determined by $P_{k}^{\phi_{k,n_{k}}}\big(a_{k,n_{k}-1}\big)$
and $P_{k}^{\phi_{k,n_{k}}}\big(\lambda_{k,1}^{\phi_{k,n_{k}},min}\big)$,
as shown in Fig.~\ref{fig:schematic_for_optimal_phases}.
The former is an increasing function with respect to $\phi_{k,n_{k}}$
and increases to 100\% when $\phi_{k,n_{k}}\to\phi_{k,n_{k}-1}$.
While, the latter monotonically decreases and asymptotically approaches
100\% when $\phi_{k,n_{k}}\to\phi_{k}^{min}$. Hence, according to
the intermediate value theorem, there exists a $\phi_{k,n_{k}}\in\big(\phi_{k}^{min},\phi_{k,n_{k}-1}\big)$
such that Eq.~(\ref{eq:proof_opt_phases_keq1_phi_nk_1}) holds.
At this time, the minimum success probability reaches the maximum,
denoted by $Q_{k}^{\phi_{k,n_{k}-1}}$.
Here, we define $Q_{k}^{\phi}$ to be the largest minimum success
probability on $\big[\lambda_{k,1}^{\phi,max},\lambda_{k-1,1}^{\pi,max}\big)$
with $\phi\in\big(\phi_{k}^{min},\pi\big]$. As $\phi$
grows, $\lambda_{k,1}^{\phi,max}=\frac{1-\cos\frac{\pi}{2k+1}}{1-\cos\phi}$
decreases and range $\big[\lambda_{k,1}^{\phi,max},\lambda_{k-1,1}^{\pi,max}\big)$
extends, then it follows that $Q_{k}^{\phi}$ monotonically decreases with
respect to $\phi$.

Next, we show that for any $\phi_{k,n_{k}-2}$,
the optimal phases condition on $\big[\lambda_{k,1}^{\phi_{k,n_{k}-2},max},\lambda_{k-1,1}^{\pi,max}\big)$
is
\begin{equation}
P_{k}^{\phi_{k,n_{k}-1}}\big(a_{k,n_{k}-2}\big)=Q_{k}^{\phi_{k,n_{k}-1}}.
\label{eq:proof_opt_phases_keq1_phi_nk_2}
\end{equation}
This is because, for a given $\phi_{k,n_{k}-2}$,
the minimum success probability on $\big[\lambda_{k,1}^{\phi_{k,n_{k}-2},max},\lambda_{k-1,1}^{\pi,max}\big)$
is determined by
$P_{k}^{\phi_{k,n_{k}-1}}\left(a_{k,n_{k}-2}\right)$
and $Q_{k}^{\phi_{k,n_{k}-1}}$,
as shown in Fig.~\ref{fig:schematic_for_optimal_phases}.
The former is an increasing function
with respect to $\phi_{k,n_{k}-1}$ and increases to 100\% when $\phi_{k,n_{k}-1}\to\phi_{k,n_{k}-2}$.
While, the latter monotonically decreases and asymptotically approaches
100\% when $\phi_{k,n_{k}-1}\to\phi_{k}^{min}$. Hence, according
to the intermediate value theorem, there exists a $\phi_{k,n_{k}-1}\in\big(\phi_{k}^{min},\phi_{k,n_{k}-2}\big)$
such that Eq.~(\ref{eq:proof_opt_phases_keq1_phi_nk_2}) holds.
At this time, the minimum success probability reaches the maximum,
denoted by $Q_{k}^{\phi_{k,n_{k}-2}}$.

In a similar way as shown before, we can obtain for any $\phi_{k,m}$
($1\le m\le n_{k}-2$), the optimal phases condition on $\big[\lambda_{k,1}^{\phi_{k,m},max},\lambda_{k-1,1}^{\pi,max}\big)$
is
\begin{equation}
P_{k}^{\phi_{k,m+1}}\left(a_{k,m}\right)=Q_{k}^{\phi_{k,m+1}}.\label{eq:proof_opt_phases_keq1_phi_m}
\end{equation}
In this case, the corresponding maximum of the minimum success probability
is denoted by $Q_{k}^{\phi_{k,m}}$.

In addition, it can be found
that for any $\phi_{k,1}$, the optimal phases condition on $\big[\lambda_{k,1}^{\pi,max},\lambda_{k-1,1}^{\pi,max}\big)$
is
\begin{equation}
P_{k}^{\phi_{k,1}}\big(\lambda_{k,1}^{\pi,max}\big)=Q_{k}^{\phi_{k,1}}.
\label{eq:proof_opt_phases_keq1_phi_1}
\end{equation}
This is because, for a given $\phi_{k,1}$, the minimum success probability
on $\big[\lambda_{k,1}^{\pi,max},\lambda_{k-1,1}^{\pi,max}\big)$
is determined by $P_{k}^{\phi_{k,1}}\big(\lambda_{k,1}^{\pi,max}\big)$
and $Q_{k}^{\phi_{k,1}}$. The former is an increasing function with
respect to $\phi_{k,1}$ and increases to 100\% when $\phi_{k,1}\to\pi$.
While, the latter monotonically decreases and asymptotically approaches
100\% when $\phi_{k,1}\to\phi_{k}^{min}$. Hence, according to the
intermediate value theorem, there exists a $\phi_{k,1}\in\big(\phi_{k}^{min},\pi\big)$
such that Eq.~(\ref{eq:proof_opt_phases_keq1_phi_1}) holds.
At this time, the minimum success probability reaches the maximum,
denoted by $Q_{k}^{\pi}$. Finally, combining Eqs.~(\ref{eq:proof_opt_phases_keq1_phi_nk_1},\ref{eq:proof_opt_phases_keq1_phi_m},\ref{eq:proof_opt_phases_keq1_phi_1}),
it is straightforward to see in the case of $k=1$, Eq.~(\ref{eq:optimal_phases_conditions})
is indeed the optimal phases condition.

In the case of $k\ge2$, from Property \ref{pro:extreme_properties_of_P_k_phi}, it follows that for any
$\phi_{k,n_{k}-1}$, the optimal phases condition on $\big[\lambda_{k,1}^{\phi_{k,n_{k}-1},max},\lambda_{k-1,1}^{\pi,max}\big)$
is
\begin{equation}
P_{k}^{\phi_{k,n_{k}}}\big(a_{k,n_{k}-1}\big)=P_{k}^{\phi_{k,n_{k}}}\big(\lambda_{k-1,1}^{\pi,max}\big).
\label{eq:proof_opt_phases_kge2_phi_nk_1}
\end{equation}
This is because, for a given $\phi_{k,n_{k}-1}$, the minimum success
probability
on $\big[\lambda_{k,1}^{\phi_{k,n_{k}-1},max},\lambda_{k-1,1}^{\pi,max}\big)$
is determined by $P_{k}^{\phi_{k,n_{k}}}\big(a_{k,n_{k}-1}\big)$
and $P_{k}^{\phi_{k,n_{k}}}\big(\lambda_{k-1,1}^{\pi,max}\big)$,
where $P_{k}^{\phi_{k,n_{k}}}\big(\lambda_{k-1,1}^{\pi,max}\big)$
similar to $P_{k}^{\phi_{k,n_{k}}}\big(\lambda_{k,1}^{\phi_{k,n_{k}},min}\big)$
in the case of $k=1$, monotonically decreases with respect to $\phi_{k,n_{k}}$
and asymptotically approaches 100\% when $\phi_{k,n_{k}}\to\phi_{k}^{min}$.
Hence, when Eq.~(\ref{eq:proof_opt_phases_kge2_phi_nk_1}) holds,
the minimum success probability reaches the maximum, denoted by $Q_{k}^{\phi_{k,n_{k}-1}}$.

Then, by the same method as employed in the case of $k=1$, it is
easy to show that the optimal phases conditions on $\big[\lambda_{k,1}^{\phi_{k,m},max},\lambda_{k-1,1}^{\pi,max}\big)$
($1\le m\le n_{k}-2$) and $\big[\lambda_{k,1}^{\pi,max},\lambda_{k-1,1}^{\pi,max}\big)$
are Eq.~(\ref{eq:proof_opt_phases_keq1_phi_m}) and Eq.~(\ref{eq:proof_opt_phases_keq1_phi_1}),
respectively. Finally, combining Eqs.~(\ref{eq:proof_opt_phases_keq1_phi_m}), (\ref{eq:proof_opt_phases_keq1_phi_1}) and (\ref{eq:proof_opt_phases_kge2_phi_nk_1}),
we can obtain in the case of $k=2$, Eq.~(\ref{eq:optimal_phases_conditions})
is also indeed the optimal phases condition. $\hfill\blacksquare$

\section{Proof of the properties of $Q_{k}^{\pi}$ with respect to $n_{k}$
on $\varLambda_{k}$ of Property \ref{pro:properties_of_Q_k_pi} \label{sub:properties_of_Q_k_pi}}

\begin{figure}[tb]
\begin{centering}
\includegraphics[width=7.5cm]{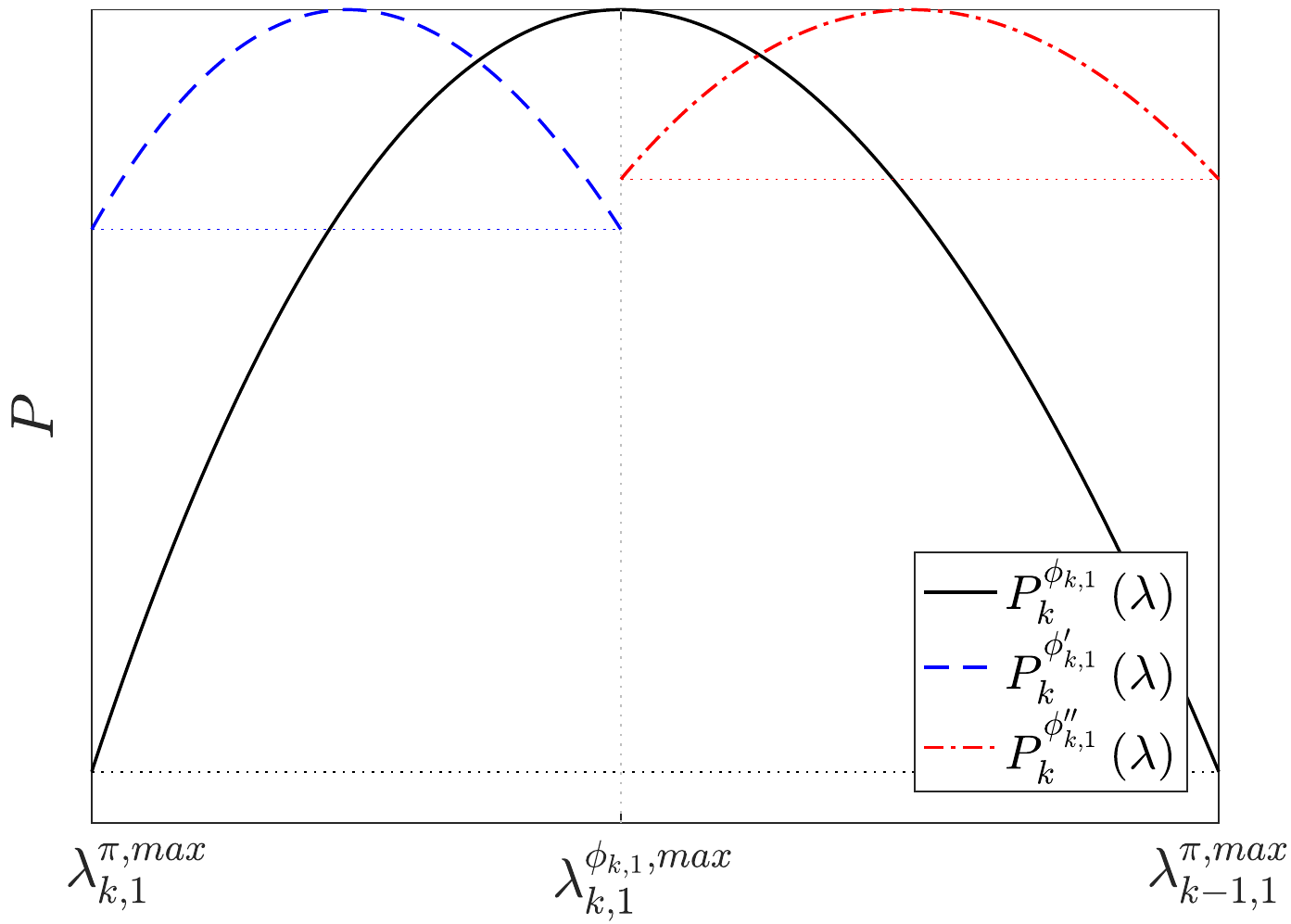}
\par\end{centering}

\protect\caption{
The schematic of success probability $P$
as a function of the fraction of target items $\lambda$ for $n_{k}=1$ and $k\ge2$.
The black solid, blue dashed, and
red dashed-dotted curves correspond to $\phi_{k,1}$, $\phi'_{k,1}$ and $\phi''_{k,1}$, respectively, which satisfy Eqs.~(\ref{eq:opt_phases_nk_eq_1}), (\ref{eq:opt_phases_nk_eq_1_left}) and (\ref{eq:opt_phases_nk_eq_1_right}), separately.
\label{fig:schematic_for_optimal_number_of_phases}}
\end{figure}

(1) The property that $Q_{k}^{\pi}$ increases as $n_{k}$ grows,
will be proven if we can show $Q_{k}^{\pi}\big(n_{k}+1\big)>Q_{k}^{\pi}\big(n_{k}\big)$
for any $n_{k}\geq1$.
In the case of $n_{k}=1$, according to Eq.~(\ref{eq:optimal_phases_conditions}),
the optimal phases condition is given as
\begin{equation}
P_{k}^{\phi_{k,1}}\big(\lambda_{k,1}^{\pi,max}\big)=P_{k}^{\phi_{k,1}}\big(\lambda_{key}\big)\equiv Q_{k}^{\pi}\big(n_{k}\big),\label{eq:opt_phases_nk_eq_1}
\end{equation}
where $\lambda_{k,1}^{\phi,max}$ and $\lambda_{key}$ are defined by Eqs.~(\ref{eq:lambda_k_1_phi_max}) and (\ref{eq:optimal_phases_conditions_lam_key}), respectively.
Without loss of generality, under condition Eq.~(\ref{eq:opt_phases_nk_eq_1}), Figure~\ref{fig:schematic_for_optimal_number_of_phases}
plots the schematic of success probability versus the fraction of target items for $k\ge2$.
On the one hand, due to
\begin{equation}
P_{k}^{\phi_{k,1}}\big(\lambda_{k,1}^{\pi,max}\big)<P_{k}^{\phi_{k,1}}\big(\lambda_{k,1}^{\phi_{k,1},max}\big)=100\%,
\end{equation}
we can see that $\phi_{k,1}$ is not the optimal phase to maximize
the minimum success probability on $\big[\lambda_{k,1}^{\pi,max},\lambda_{k,1}^{\phi_{k,1},max}\big)$
of the algorithm using one phase. In other words, as shown in Fig.~\ref{fig:schematic_for_optimal_number_of_phases}, there exists a
$\phi'_{k,1}\in\big(\phi_{k,1},\pi\big]$ such that
\begin{equation}
P_{k}^{\phi'_{k,1}}\big(\lambda_{k,1}^{\pi,max}\big)=
P_{k}^{\phi'_{k,1}}\big(\lambda_{k,1}^{\phi_{k,1},max}\big)
>Q_{k}^{\pi}\big(n_{k}\big).\label{eq:opt_phases_nk_eq_1_left}
\end{equation}
On the other hand, from
\begin{equation}
P_{k}^{\phi_{k,1}}\big(\lambda_{k,1}^{\phi_{k,1},max}\big)=
100\%>P_{k}^{\phi_{k,1}}\big(\lambda_{key}\big),
\end{equation}
it follows that $\phi_{k,1}$ is neither the optimal phase on $\big[\lambda_{k,1}^{\phi_{k,1},max},\lambda_{k-1,1}^{\pi,max}\big)$
of the algorithm with a single phase.
Namely, as illustrated in Fig.~\ref{fig:schematic_for_optimal_number_of_phases}, there exists a $\phi''_{k,1}\in\big(\phi_{k}^{min},\phi_{k,1}\big)$,
such that
\begin{equation}
P_{k}^{\phi''_{k,1}}\big(\lambda_{k,1}^{\phi_{k,1},max}\big)=
P_{k}^{\phi''_{k,1}}\big(\lambda''_{key}\big)>Q_{k}^{\pi}\big(n_{k}\big),
\label{eq:opt_phases_nk_eq_1_right}
\end{equation}
where
\begin{equation}
\lambda''_{key}=\begin{cases}
\lambda_{k,1}^{\phi''_{k,1},min}, & {\rm if}\thinspace k=1,\\
\lambda_{k-1,1}^{\pi,max}, & {\rm if}\thinspace k\ge2.
\end{cases}
\end{equation}

Then, using $\phi'_{k,1}$ and $\phi''_{k,1}$ yields a minimum success
probability on $\varLambda_{k}$, which is greater than $Q_{k}^{\pi}\big(n_{k}\big)$.
Furthermore, under the optimal phases condition, the minimum success
probability using two phases will be greater. Thus,
$Q_{k}^{\pi}\big(n_{k}+1\big)>Q_{k}^{\pi}\big(n_{k}\big)$ is
confirmed for $n_{k}=1$.

In the case of $n_{k}\ge2$, according to Eq.~(\ref{eq:optimal_phases_conditions}),
the optimal phases condition is given as
\begin{eqnarray}
P_{k}^{\phi_{k,1}}\big(\lambda_{k,1}^{\pi,max}\big)
&=&P_{k}^{\phi_{k,2}}\big(a_{k,1}\big)
=\cdots=P_{k}^{\phi_{k,n_{k}}}\big(a_{k,n_{k}-1}\big)\nonumber \\
&=&P_{k}^{\phi_{k,n_{k}}}\big(\lambda_{key}\big)
\equiv Q_{k}^{\pi}\big(n_{k}\big).\label{eq:opt_phases_nk_eq_m}
\end{eqnarray}
On the one hand, due to
\begin{equation}
P_{k}^{\phi_{k,1}}\big(\lambda_{k,1}^{\pi,max}\big)<P_{k}^{\phi_{k,n_{k}}}\big(\lambda_{k,1}^{\phi_{k,n_{k}},max}\big)=100\%,
\end{equation}
we can find that $\phi_{k,1},\phi_{k,2},\cdots,\phi_{k,n_{k}}$ are not
the optimal phases on $\big[\lambda_{k,1}^{\pi,max},\lambda_{k,1}^{\phi_{k,n_{k}},max}\big)$
of the algorithm using $n_{k}$ phases.
In other words, there exist
$\phi'_{k,1},\phi'_{k,2},\cdots,\phi'_{k,n_{k}}\in\big(\phi_{k,n_{k}},\pi\big]$
such that
\begin{eqnarray}
P_{k}^{\phi'_{k,1}}\big(\lambda_{k,1}^{\pi,max}\big)
&=&P_{k}^{\phi'_{k,2}}\big(a'_{k,1}\big)=\cdots
=P_{k}^{\phi'_{k,n_{k}}}\big(a'_{k,n_{k}-1}\big)\nonumber \\
&=&P_{k}^{\phi'_{k,n_{k}}}\big(\lambda_{k,1}^{\phi_{k,n_{k}},max}\big)
>Q_{k}^{\pi}\big(n_{k}\big),
\end{eqnarray}
where $a'_{k,j}$ denotes the intersection point of
$P_{k}^{\phi'_{k,j}}\left(\lambda\right)$ and $P_{k}^{\phi'_{k,j+1}}\left(\lambda\right)$, $1\le j\le n_{k}-1$.
On the other hand, from
\begin{equation}
P_{k}^{\phi_{k,n_{k}}}\big(\lambda_{k,1}^{\phi_{k,n_{k}},max}\big)=100\%>P_{k}^{\phi_{k,n_{k}}}\big(\lambda_{key}\big),
\end{equation}
it follows
that $\phi_{k,n_{k}}$ is neither the optimal phase on $\big[\lambda_{k,1}^{\phi_{k,n_{k}},max},\lambda_{k-1,1}^{\pi,max}\big)$
of the algorithm with a single phase. Namely, there exists
a $\phi''_{k,n_{k}}\in\big(\phi_{k}^{min},\phi_{k,n_{k}}\big)$
such that
\begin{equation}
P_{k}^{\phi''_{k,n_{k}}}\big(\lambda_{k,1}^{\phi_{k,n_{k}},max}\big)=P_{k}^{\phi''_{k,n_{k}}}\big(\lambda''_{key}\big)>Q_{k}^{\pi}\big(n_{k}\big),
\end{equation}
where
\begin{equation}
\lambda''_{key}=\begin{cases}
\lambda_{k,1}^{\phi''_{k,n_{k}},min}, & {\rm if}\thinspace k=1,\\
\lambda_{k-1,1}^{\pi,max}, & {\rm if}\thinspace k\ge2.
\end{cases}
\end{equation}

Then, using $\phi'_{k,1},\phi'_{k,2},\cdots,\phi'_{k,n_{k}}$ and
$\phi''_{k,n_{k}}$ will yield a minimum success probability on $\varLambda_{k}$,
greater than $Q_{k}^{\pi}\big(n_{k}\big)$. Furthermore,
under the optimal phases condition, the minimum success probability
of the algorithm using $n_{k}+1$ phases will be greater. Thus, $Q_{k}^{\pi}\big(n_{k}+1\big)>Q_{k}^{\pi}\big(n_{k}\big)$
is confirmed for $n_{k}\ge2$. At this point, the property that $Q_{k}^{\pi}$
increases as $n_{k}$ grows is proven.

(2) First, we can equally divide $\varLambda_{k}$ into $n_{k}$ smaller
ranges, denoted by $\varLambda_{k,1}$, $\varLambda_{k,2}$, $\cdots$,
$\varLambda_{k,n_{k}}$. For each $\varLambda_{k,m}$,
there exists a phase $\phi_{k,m}$ such that $\lambda_{k,1}^{\phi_{k,m},max}\in\varLambda_{k,m}$, $1\le m\le n_{k}$.
When $n_{k}\to\infty$, the length of $\varLambda_{k,m}$, i.e.,
\begin{equation}
\big(\lambda_{k-1,1}^{\pi,max}-\lambda_{k,1}^{\pi,max}\big)/n_{k}\to0,
\end{equation}
which yields that for any $\lambda\in\varLambda_{k,m}$,
\begin{equation}
P_{k}^{\phi_{k,m}}\big(\lambda\big)\to P_{k}^{\phi_{k,m}}\big(\lambda_{k,1}^{\phi_{k,m},max}\big)=100\%.
\end{equation}
Furthermore, under the optimal
phase condition, the minimum success probability of the algorithm
using $n_{k}$ phases will be greater. Consequently, it is straightforward
to show that $Q_{k}^{\pi}\to100\%$ when $n_{k}\to\infty$. $\hfill\blacksquare$


%

\end{document}